\RequirePackage[prologue,table,dvipsnames]{xcolor}
\RequirePackage{snapshot}
\documentclass[acmlarge,timestamp,letterpaper]{acmart}

\usepackage[many]{tcolorbox}
\usepackage{varwidth}
\usepackage{environ}
\usepackage{xparse}
\usepackage{microtype}
\usepackage{tikz}
\usepackage{calc}

\newlength{\somelengthA}
\newlength{\somelengthB}
\newlength{\somelengthC}

\newcommand\clock[3]{%
\begin{tikzpicture}[line cap=round,line width=3pt]
\filldraw [fill=Goldenrod!20,thin] (0,0) circle (8cm);
\draw [gray,thin] (0,0) circle (5.2cm);
\draw [gray,thin] (0,0) circle (2.8cm);
\draw [gray,thin] (0,0) circle (.42cm);
\node (0,0) {\includegraphics[width=1.2em]{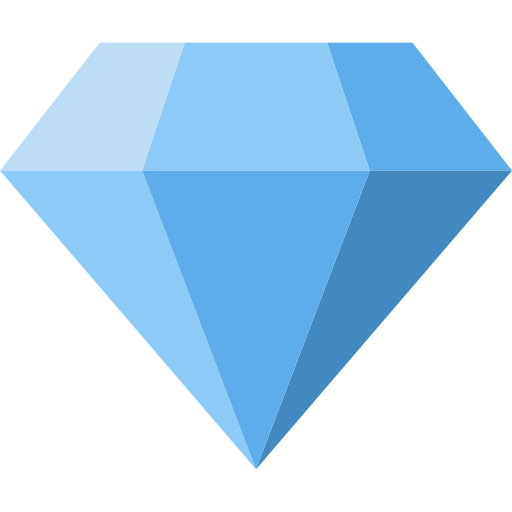}};
\foreach \angle / \label / \rotate in
{
  33/PAR→/-55,
  147/DPL→/55,
  270/←CLA/0
}
{
\draw[line width=1pt] (\angle:8cm) -- (\angle:8.2cm);
\draw (\angle:8.6cm) node[rotate=\rotate]{\textsf{\label}};
}
\foreach \angleA / \labelA in
{
  23/{Dérive Comix},
  13/{Share Back},
  115/{Meaning Map},
  124/{Components},
  127/{Language}, 
  130/{Pattern},
  140/{Going Meta},
  265/{Reinfuse},
  262/{Expertise},
  242/{Roles},
  245/{Functional},
  224/{Control},
  227/{Participant},
  230/{Increase}
}
{
\settowidth{\somelengthA}{\labelA}
\draw (\angleA:6.6cm) node[rotate=\angleA,transform canvas={xshift=\somelengthA}]{\textsf{\labelA}};
}
\foreach \angleB / \labelB / \corr in
{
  -0/{Context}/0,
  -4/{Setting}/-3,
  109/{Facilitator}/-2,
  104/{Roles}/-7
}
{
\settowidth{\somelengthB}{\labelB}
\draw (\angleB:4cm) node[rotate=\angleB-\corr,transform canvas={xshift=\somelengthB}]{\textsf{\labelB}};
}
\foreach \angleB / \labelB / \corr in
{
  -13/{Time Traveler}/0,
  -23/{Reflector}/0,
  85/{Linker}/0,
  95/{Analyst}/0,
  220/{Wrinkler}/0,
  207/{Stepper}/0
}
{
\settowidth{\somelengthB}{\labelB}
\draw (\angleB:4cm) node[rotate=\angleB-\corr,transform canvas={xshift=\somelengthB}]{\bfseries\scshape{\labelB}};
}
\foreach \angleC / \labelC / \corr in
{
  -33/{Do Your}/0,
  -45/{Research}/10,
  190/{Conversations}/0,
  200/{Structure}/-10,
  65/{Begins Now}/0,
  80/{The Future}/-14
}
{
\settowidth{\somelengthC}{\labelC}
\draw (\angleC:1.65cm) node[rotate=\angleC+\corr,transform canvas={xshift=\somelengthC}]{\textsf{\labelC}};
}
\end{tikzpicture}%
}

\usepackage{slantsc}

\setcopyright{rightsretained}
\copyrightyear{2023}

\acmPrice{15.00}
\acmISBN{978-1-9416-5219-0/18/06}

\usepackage{graphicx}
\usepackage{grffile}
\usepackage{longtable}
\usepackage{wrapfig}
\usepackage{rotating}
\usepackage[normalem]{ulem}
\usepackage{textcomp}
\usepackage{capt-of}
\usepackage{hyperref}

\usepackage{mdframed}
\usepackage{afterpage}

\usepackage{booktabs}

\usepackage[inline]{enumitem}

\usepackage[pagewise]{lineno}

\usepackage{natbib}
\usepackage{float}
\usepackage{tikz}

\usepackage{placeins}
\newcommand{\candidate}[1]{{\scitshape #1}}
\usepackage{bold-extra}
\DeclareRobustCommand{\scitshape}{\fontshape{\scitdefault}\selectfont}

\usepackage{enotez}
\renewcommand{\endnote}[1]{}

\newcommand{\markbf}[1]{\textsuperscript{\textbf{#1}}}
\setenotez{counter-format = alph, mark-cs = \markbf}
\DeclareInstance{enotez-list}{sverre}{paragraph}{heading={},notes-sep=\baselineskip,format=\normalsize\normalfont\raggedright\leftskip1.8em,number=\makebox[0pt][r]{#1.\ }\ignorespaces,}
\usepackage{epigraph}
\date{}
\title{Patterns of Patterns II: Discourse on Implementation}
\hypersetup{
 pdfauthor={Joseph Corneli et al.},
 pdftitle={Patterns of Patterns II},
 pdfkeywords={},
 pdfsubject={},
 pdfcreator={Emacs 30.0.50 (Org mode 9.6.1)},
 pdflang={English}}

\begin{document}

\title{Patterns of Patterns II}

\author{Joseph Corneli}
\authornote{Corresponding author, jcorneli@brookes.ac.uk.}
\email{jcorneli@brookes.ac.uk}
\orcid{1234-5678-9012}
\affiliation{%
  \institution{Oxford Brookes University}
  \streetaddress{Gipsy Lane}
  \city{Oxford}
  \country{UK}
  \postcode{OX3 0BP}
}
\affiliation{%
  \institution{Hyperreal Enterprises Ltd}
  \streetaddress{272 Bath Street}
  \city{Glasgow}
  \country{UK}
  \postcode{G2 4JR}}

\author{Noorah Alhasan}
\author{Leo Vivier}
\author{Alex Murphy}
\author{Raymond S. Puzio}
\email{rsp@hyperreal.enterprises}
\affiliation{%
  \institution{Hyperreal Enterprises Ltd}
  \streetaddress{272 Bath Street}
  \city{Glasgow}
  \country{UK}
  \postcode{G2 4JR}}

\author{Abby Tabor}
\affiliation{%
  \institution{University of the West of England}
  \streetaddress{Faculty of Health and Applied Sciences (HAS), Frenchay Campus, Coldharbour Lane}
  \city{Bristol}
  \state{England}
  \country{UK}
  \postcode{BS16 1QY}}
\email{abby.tabor@uwe.ac.uk}



\author{Sridevi Ayloo}
\affiliation{%
  \institution{New York City College of Technology}
  \streetaddress{300 Jay St}
  \city{Brooklyn}
  \postcode{11201}
  \country{USA}
}
\email{SAyloo@CityTech.Cuny.Edu}

\author{Mary Tedeschi}
\author{Manvinder Singh}
\author{Kajol Khetan}
\affiliation{%
  \institution{Baruch College}
 \streetaddress{PO Box 802738}
 \city{New York}
 \state{NY}
  \country{USA}
  \postcode{60680}}
\email{mtedeschi@pace.edu}

\author{Charles J. Danoff}
\affiliation{%
  \institution{Mr Danoff’s Teaching Laboratory}
 \streetaddress{PO Box 802738}
 \city{Chicago}
 \state{IL}
  \country{USA}
  \postcode{60680}}
\email{contact@mr.danoff.org}

\renewcommand{\shortauthors}{Corneli et al.}

\begin{abstract}
Our earlier paper “Patterns of Patterns” combined three techniques from training, futures studies, and design in a design pattern called PLACARD that helps groups of people work together effectively.  We used that pattern in five hands-on workshop case studies which took place at various locations in the US and the UK.  This experience report documents what we learned, including the way our thinking about PLACARD evolved, together with additional patterns our work generated.  We evaluate the reproducibility of our methods and results, and consider the broader economic implications of this way of working. We discuss implications of our prototyping work for the design of future platforms, drawing connections with recent developments in cognitive science and artificial intelligence.  This positions our patterns of patterns as a toolkit for the design and governance of systems that combine social dynamics with technical components.

\end{abstract}

\ccsdesc[500]{Social and professional topics}
\ccsdesc[300]{Software and its engineering~Designing software}
\ccsdesc[300]{Software and its engineering~Open source model}
\ccsdesc[300]{Applied computing~Operations research}
\ccsdesc[100]{Computing methodologies~Modeling and simulation}

\keywords{Design Patterns, Pattern Languages, Action Reviews, Futures
Studies, Causal Layered Analysis, Free Software, Peeragogy,
Artificial Intelligence, Anticipation, Doughnut Economics, Socio-Technical Systems}


\maketitle


\section{Introduction}
\label{sec:org195e8e3}
\label{Introduction}

This paper is primarily an experience report describing a series of workshops exploring pattern-based methods for co-creation.
Patterns served a design function in setting up the workshops, and were discussed as contents of the workshops.
Furthermore, patterns were engaged at all stages of development.
Our aim is to understand and express the potential that is implied by
design pattern language methods.
The paper itself has the following structure.
We begin with a brief recapitulation of the prequel, “Patterns of
  Patterns” \cite{patterns-of-patterns-i}.
We then motivate our approach to design pattern methods, and provide
  an overview of the patterns we identified across five workshops,
  which become case studies in our treatment here.  In brief, we show
  how our methods became more reliable as repeatable patterns, with
  implications across increasingly broad contexts of application.
We then discuss connections to related work, positioning this paper
  as a contribution to contemporary discourse about openness and
  reproducibility in qualitative research.
We conclude with our assessment of how these design patterns
  could inform the design of future platforms for co-creation.
For concision, details of the patterns themselves are presented
  in an Appendix.


\section{Background} \label{sec:Background}

In “Patterns of Patterns” \cite{patterns-of-patterns-i}, we introduced
a synthesis of methods that operationalize social intelligence.  The
particular methods we outlined were certainly not the only way to
implement the necessary system features.  What drew our attention is
that each of the methods we selected comes with a framework or
template; each of the methods is, essentially, a design pattern.

\begin{itemize}
\item Project Action Review (PAR): \emph{a set of five review questions to explore at a project checkpoint}.
\item Causal Layered Analysis (CLA): \emph{a set of four “layers” that can
be used to unpack a topic of interest.}
\item Design Pattern Languages (DPL): \emph{each constituent pattern follows a three-part “Context/Problem/Solution” template.}
\end{itemize}


The Project Action Review adapts the precursor After Action Review
(AAR) to horizontally-managed group work.  The AAR has seen
widespread use in military, business, and healthcare settings, and has
proved to be an effective training approach in task environments
characterized by a combination of high complexity and ambiguity
\cite{Keiser2021,Keiser2022}.  Causal Layered Analysis has been used
for over thirty years by members of the futuring community to scaffold
change-making initiatives by understanding and changing the
core metaphor or “myth” that underpins a given social situation, along
with its relationship to other levels of social organization \cite{cla3}.  Design Pattern Languages are
comprised of \emph{design patterns}, which are methods that (\emph{1})
describe a context (\emph{2}) in which a recognizable problem occurs,
together with (\emph{3}) a solution strategy that should resolve the
problem, possibly with reference to further patterns that provide
needed detail.  We called the overall method that combines the three
methods PAR, CLA, and DPL the “PLACARD” pattern.  We elaborated our
description of PLACARD with case studies, and used our analysis of
these to outline four potential futures for the development of
pattern methods.

Table \ref{four-domains} arranges these four futures schematically in
relationship to the four economic domains they are most closely
associated with: the \emph{household}, the \emph{state}, the
\emph{market}, and the \emph{commons}
\cite[p.~71]{raworth2017doughnut}.  The potential for design to
interact fruitfully with these domains has been noted by
\citet{BOEHNERT2018355}.  We observe that patterns only reach their
potential in these domains if they are accompanied by certain
\emph{freedoms}, familiar from the Free Software
movement.\footnote{\url{https://www.gnu.org/philosophy/free-sw.html}}
For example, if patterns were to become computational, they could be
used like any other computational service (or other household device).
Legal protections would need to be added to ensure access to source
code (present for Wikipedia, absent for Netflix).  This is important
because patterns need to be adapted to local circumstances before they
can be applied.  However, benefits will only accrue locally or to
private individuals, unless further design intelligence is applied to
develop and share patterns that empower everyone, and which take
account of our role within broader ecosystems.


\begin{table}
\begin{center}
\begin{tabular}{llp{2.2in}}
Economic Domain & Potential Future for Design Patterns    & Related Software Freedom\\[0pt]
\hline
\textbf{household}       & patterns become computational         & “The freedom to run the program as you wish, for any purpose (freedom 0).”\\[0pt]
\textbf{state}           & free/libre/open source licensing      & “The freedom to study how the program works, and change it so it does your computing as you wish (freedom 1).”\\[0pt]
\textbf{market}          & empower individuals and communities   & “The freedom to redistribute copies so you can help others (freedom 2).”\\[0pt]
\textbf{commons}         & widespread economic empowerment       & “The freedom to distribute copies of your modified versions to others (freedom 3).”\\[0pt]
\end{tabular}
\end{center}
\caption{Analogy between Raworth’s economic domains, our potential pattern futures, and Stallman’s four software freedoms\label{four-domains}}
\end{table}

\section{Methodology}
\label{sec:org134acbb}
\label{methods}


The workshop methodology that we presently employ traces its origins
back to a presentation at the 2019 Anticipation conference.\footnote{A
fictional peeragogical anticipatory learning exploration, 11 October
2019, Paper Session 13,
\url{http://anticipationconference.org/wp-content/uploads/2019/10/Anticipation_2019_paper_113.pdf}}
Our goal at the time was to provide a rapid introduction to our work
on peeragogy.\footnote{\url{https://peeragogy.org}} We wanted to use a
format in which the audience would participate actively, since we
thought that would better embody the spirit of peeragogy than a
lecture.  Thus, we wrote our presentation in the form of a dramatic
dialogue and asked attendees to pick parts and perform the dialogue.
While the attendees appreciated the opportunity to participate, one of
them remarked that use of a pre-determined script felt confining.

In response to these comments, we developed a more open-ended approach
and refined it in pilot workshops.  Instead of handing participants a
finished script, we provided them with design patterns and functional
roles as a framework within which to improvise their own dialogue.
Moreover, we made it clear that these roles and patterns were are only
meant as a starting point and encouraged them to reinterpret and
modify the material they were given, and to develop new patterns in
the course of the workshop.


Our starting assumptions were that a short training in PLACARD methods
and the patterns that operationalize those methods would help
participants who were not familiar with Peeragogy
\cite{corneli2015a,peeragogy-handbook}—and who, indeed, had never
previously met each other—learn together and collaborate effectively
in the workshop setting.  We also expected that these workshops would
support a form of distributed collaboration, across workshop contexts
and topics, as we gathered themes and insights from people who would
never meet at all.
During the course of an individual workshop and over the longer time-scale of multiple workshops, methods and patterns evolved.  Their
interpretation changed in accordance with community understanding, and
new patterns emerged from repeated experiences.  A representative
selection of the patterns can be found in an \hyperref[catalogue]{Appendix},
and Table \ref{summary} presents an overview.  The
high-level subdivisions in this table mirror the structure of PLACARD:
“Identifying themes”, “Organizing structure” and “Making it
actionable”.  In some of the workshops, the design patterns were
shared with attendees in the form of pattern cards which could be used
as in-workshop manipulatives.  In others, the patterns were primarily
used to design and analyze the workshop, and stayed more or less
behind the scenes.  Our case study descriptions clarify those details.

We assess each of the case studies using a structured framework we use to
reflect on which aspects of the case study are reproducible: methods, results,
interpretation \cite{Goodman2016}.  These reflect PLACARD's three
dimensions---\emph{methods reproducibility} pertains to the motor task of replicating an experimental set-up form a published description; \emph{results reproducibility} pertains to the sensory task of collecting data from a potential replication, and comparing results of a previous experiment; and \emph{interpretation reproducibility} pertains to the cognitive task of verifying the processes of analysis and inference used to draw conclusions from data.
As we draw our interpretations here, as part of our per-section evaluations, we flag the most closely related domain in Table \ref{four-domains}, to highlight the potential reach/impact of each case study.




\begin{table}[p]
\hspace{-1em}\begin{tabular}{rp{.82\textwidth}}
    PLACARD & ‘By using the PAR (or another sensory method), we are
    able to identify recurring themes.  Then, by using the CLA (or
    another cognitive method), we are able to organize these repeating
    themes in a structure that exposes the underlying trends, causes,
    and potential terminating states. With DPL (or another motor
    method) we can make what we have learned actionable.’ \cite{patterns-of-patterns-i}
\end{tabular}
\begin{center}
$\Rightarrow$\hspace{1em}\textbf{Process}: Workshop design, delivery, and analysis adds actionable detail to that proposition.
\end{center}

\vspace{.1in}
  \begin{tabular}{rll}
&\hspace{-.7in}\emph{Identifying themes:}&\\[2mm]
{[}1{]}&{\sc Dérive Comix}& ‘document what you see’ \\
&{\sc Share Back}& ‘individual groups should present key findings’\\
&\candidate{Pilot to Anticipate}& ‘anticipate the issues likely to arise in future iterations’\\
\cline{2-3}
{[}2{]}&{\sc Context Setting}& ‘describe the hoped-for outcomes’\\
&{\bfseries\scshape Time Traveler }& ‘provide historical context and
anticipate alternate futures’\\
&{\bfseries\scshape Reflector}& ‘appraise each developing scenario’\\
&\candidate{Contested Space} \includegraphics[width=1.2em]{gem}& ‘each space need not support every use equally’ \\
\cline{2-3}
{[}3{]}&{\sc Do Your Research}& ‘start doing the research in a more centralized way’\\[1em]
&\hspace{-.7in}\emph{Organizing structure:}& \\[2mm]
{[}1{]}&{\sc Meaning Map }& ‘get everyone on the same page’\\
&{\sc Pattern Language Components }& ‘build patterns piece by piece’\\
&{\sc Going Meta}& ‘explore how the project’s methods can be applied to
itself’\\
\cline{2-3}
{[}2{]}&{\sc Facilitator Roles}& ‘structure the work’\\
&{\bfseries\scshape Analyst }& ‘identify and orchestrate the dynamic network’ \\
& {\bfseries\scshape Linker}& ‘providing visualization of patterns and interconnections’ \\
&\candidate{Funding of Public Space} \includegraphics[width=1.2em]{gem}& ‘create a register of impacts’\\
\cline{2-3}
{[}3{]}&{\sc Structure Conversations}& ‘structure the discussions around shared interests’\\
&\candidate{Destructure Patterns}& ‘a less formal discussion can surface useful meanings’\\
\cline{2-3}
{[}4{]}&\candidate{Adapt Layers As Needed}& ‘layer-based analysis facilitates effective communication’\\
&\candidate{Avoiding Mistakes}& ‘navigate common project development pitfalls’\\
&\candidate{Scaling and Adaptability}& ‘aim to seamlessly adopt new advancements’\\[1em]
&\hspace{-.7in}\emph{Making it actionable:}& \\[2mm]
{[}1{]}&{\sc Reinfuse Expertise }& ‘enable richer and more complex thinking’\\
&{\sc Functional Roles}& ‘introduce different perspectives’\\
&\candidate{Increase Participant Control}& ‘participants should not remain only an audience’\\
\cline{2-3}
{[}2{]}&{\bfseries\scshape Wrinkler}& ‘what might derail or counter the proposed solution’\\
&{\bfseries\scshape Stepper}& ‘decide which actions would be most useful’\\
&\candidate{Rebalance Social Services} \includegraphics[width=1.2em]{gem}& ‘address complex local challenges’\\
\cline{2-3}
{[}3{]}&{\sc The Future Begins Now}& ‘take preliminary actions before leaving’\\
&\candidate{Structure Outputs}& ‘link intermediate artifacts into a relevant template’\\
\cline{2-3}
{[}4{]}&\candidate{Engagement and Guidance}& ‘create a collaborative learning environment’\\
  \end{tabular}
  \medskip
  \caption{A summary of our collected “Patterns of Patterns”.  Legend: Small caps is used for formalized patterns; bold is used to denote roles; italic serif font is used to denote proto-patterns.   Proto-Patterns that were created by workshop participants in Case Study 2 are further distinguished by a “\scalebox{.8}{{\normalsize \includegraphics[width=1.2em]{gem}}}” marker.  Numbers in the left-most column cross-reference the case studies ([1], [2], [3] and [4]) in which the patterns were initially developed and used.\label{summary}}
\end{table}

\FloatBarrier

\section{Case Study 1: “Going Meta” workshop at Anticipation 2022}\label{anticipation-case-study}

This workshop functioned as a more developed pilot of methods that we
had previously trialed at PLoP 2021 as “Flaws of the Cool
City”\footnote{\url{https://www.hillside.net/plop/2021/index.php?nav=program\#focusgroups}}
and at the Oxford Brookes Creative Industries Festival as “Dragon
versus monkey: A kaijū introduction to
peeragogy”.\footnote{\url{https://www.brookes.ac.uk/research/networks/creative-industries-research-and-innovation/festival-2021}}
Our central aim was to ‘workshop’ the methods with attendees.    Our
pitch was that the workshop would help attendees establish a position of
maximum leverage, exercising “Critical Anticipatory Capacities” and
using “Creativity, Innovation and New Media” (two of the conference’s
themes) to explore the future of anticipation.



\begin{figure}[h]
\begin{tabular}{c@{\hspace{1em}}c}
\includegraphics[width=.45\textwidth]{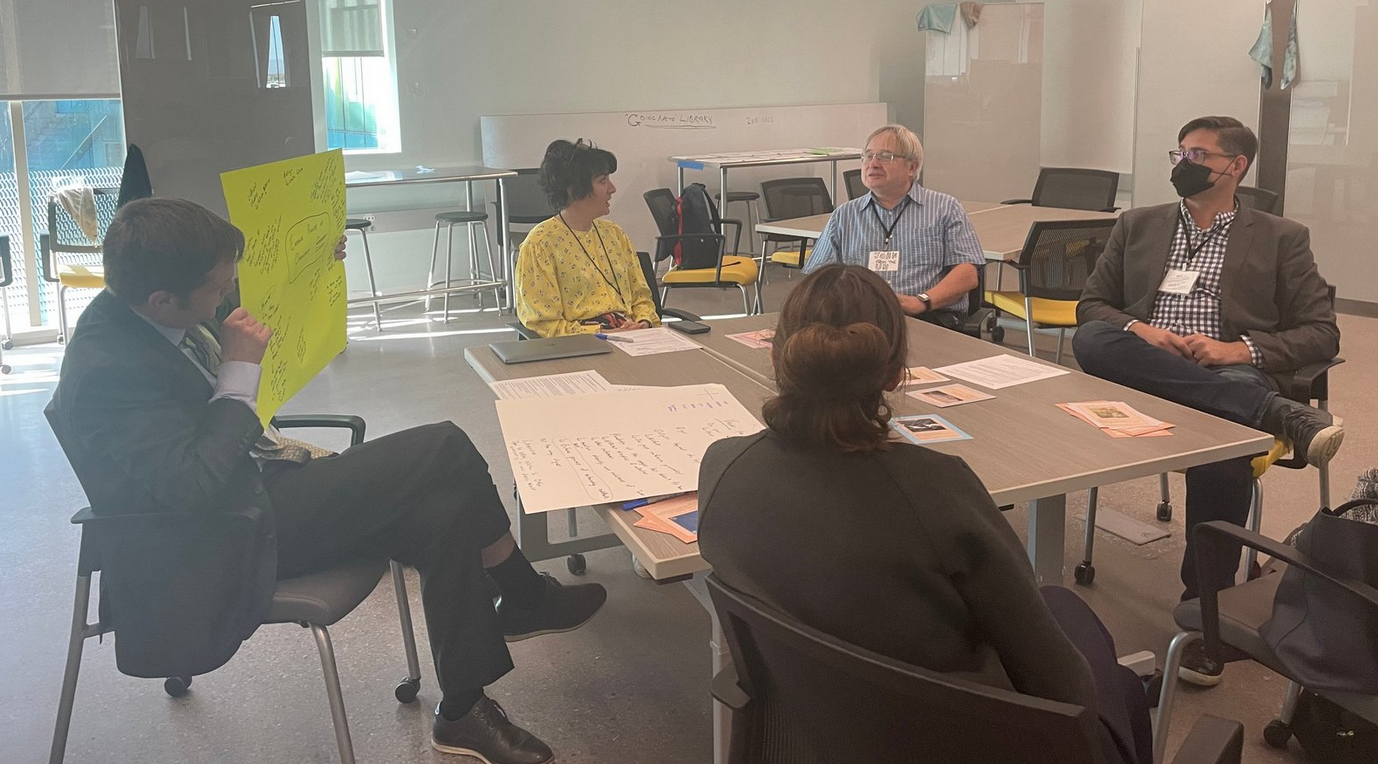}  &
\includegraphics[width=.45\textwidth]{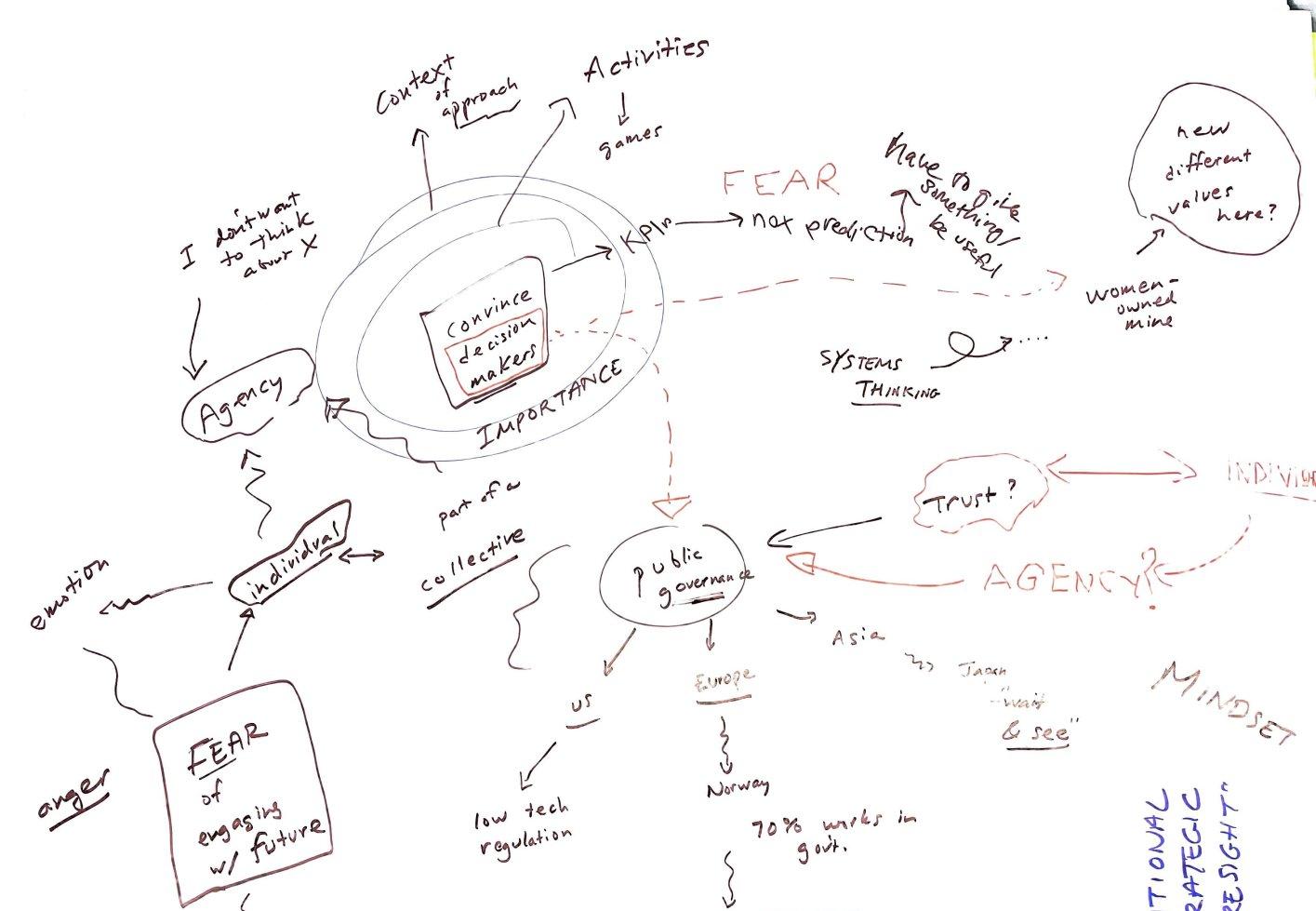}
\end{tabular}
\caption{Left: In the workshop, participants used pattern description
  cards to structure discussions (cf.~ the {\sc Pattern Language
    Components} and {\sc Functional Roles} patterns).  Right:
  Facilitators took notes and made diagrams ({\sc Meaning
    Map}).\label{anticipation-workshop}}
\end{figure}

\noindent
Figure \ref{anticipation-workshop} shows the workshop in process
(left) and details of the notes taken by one of the facilitators
(right).  Facilitators moved between facilitating small-group and
whole-group activities.  The discussion ranged over various themes,
including collective agency and how to engage more people in futures
discourse.

In the concluding Project Action Review, participants had different
responses to our request to reflect on the workshop’s activities.
Some considered the workshop to have led to good conversations, but
doubted if the manipulatives and other structure helped.  However, as mentioned in Section \ref{Introduction},  we
had observed that some participants had used the manipulatives
in creative ways—such as asking each other to pass around the cards to
better narrate the points they were making in the conversation.  Even if another conversation with these attendees would
have been good, it would have been a very different conversation.
Overall, workshop attendees had plenty of feedback for us about how we
could improve our use of the
methods;\footnote{\url{https://groups.google.com/g/peeragogy/c/V-knbZkwhB0/m/jUbw3_I9AAAJ}}
they did not, however, adopt any actions for themselves to take
forward around the points we discussed.  This was in keeping with our
primary aim for the workshop, but somewhat surprising since we were
used to Project Action Reviews helping to establish a collective sense
of direction.  We reflected that we could make that particular aim
clearer in future workshops.


\subsection{Evaluation}
\begin{description}
    \item[Methods] These or similar pattern cards can be re-used, as
      can the meeting itinerary (indeed, later case studies follow a
      similar itinerary).
    \item[Results] As discussed above, we would likely have had a good
      discussion with the participants without the manipulatives, but
      (\emph{a}) that would have had extremely low reproducibility; and, in
      particular, (\emph{b}) we would not have received feedback on the
      manipulatives had we not used and presented them.
    \item[Interpretation] The fact that we received usable feedback on
      the materials shows that they can be used in a co-design
      process.  However, the apparent resistance by attendees to
      forming an action plan based on our discussion suggested that we
      were the primary beneficiaries of this “methods workshop”.  The
      workshop was effectively an open house, and the feedback
      applied, correspondingly, at the \textbf{household} level.
\end{description}

\section{Case Study 2: Public Space for Public Health}\label{second-case-study}

This workshop was commissioned by co-author Abby Tabor as part of her
research project at the University of the West of England on
“Designing urban environments for human health: from the microbiome to
the metropolis”.  The aim was to gather attendees with an interest in
the project themes and work together to envision next
steps. Elaborations of these were developed by participants, and were
organized by facilitators using a software tool based on Org Roam and
Org Roam UI.

Inspired by our experience in the Anticipation pilot, here looked for
new methods to {\sc Increase Participant Control}.  In particular, we
generated some further articulations of the {\sc Functional Roles}
that would help with this.  The roles presented come with mnemonic
symbols based on the chess set: at the workshop, participants were
provided with additional physical manipulatives, i.e., the chess
pieces that correspond to the symbols here, along with new pattern
cards.  As mentioned previously, in the Anticipation pilot, the roles
were aligned with the {\sc Pattern Language Components}.
In the pilot, the “{\scitshape Kaij\=u Communicator}” role had
final say over
the challenges implied by the ‘HOWEVER’ keyword.\footnote{\url{https://hyperreal.enterprises/open-future/}}  Here, this role was
renamed and adapted as the “{\sc Wrinkler}”, and ascribed a
conversational perspective rather than a “role-playing persona” that
would be taken on by whomever was holding the card.  In
still-earlier pilots, the roles had been assigned more elaborate
responsibilities, and participants would receive a brief training for
the role prior to taking it on.  For example, we had made use of a
{\scitshape Designer} role, now dropped entirely, which was to be
filled after a briefing on design pattern methods and a specific
collection of design patterns.


As presented in this workshop, the roles were strictly functional and
were not the focus of role play as such.  Rather, an attendee would
fill a given role momentarily within a conversation, while remaining
fully themselves.  The roles were again outlined as pattern cards,
together with an informal verbal description.  The reader may refer to
Table \ref{mnemonic-for-manipulatives} which explains the mnemonic
meaning of the chess pieces which accompanied each role description.
The {\sc Linker} role was filled by offsite facilitators, and the {\sc
  Reflector} role was filled by onsite facilitators.  A designated
{\sc Stepper} role was only proposed after the workshop, though
participants were asked to look for “Next Steps” in putting the
patterns they were prototyping into action (in the style of
\cite{corneli2015a}).  These were elaborated using the ‘SPECIFICALLY’
keyword from the {\sc Pattern Language Components}.  This served a
core purpose in the event, intended to address the limitation from the
previous workshop.  As noted, previously, the implied “Next Steps”
were all presented in the form of feedback or advice for the workshop
facilitators, rather than as actions which the attendees could (at
least in principle) adopt and complete
themselves.\footnote{\url{https://www.eventbrite.co.uk/e/public-space-for-public-health-a-call-to-action-tickets-492522286417}}

In the spirit of {\sc Increasing Participant Control}, the {\sc
  Facilitator roles} could be distributed to participants, though
doing so in a future workshop would require spending more time on
training, and potentially also improved tooling for moderating the
flow of information within (and beyond) the workshop.

\begin{table}[h]
\begin{tabular}{llp{.5\textwidth}}
\textbf{Role} & \textbf{Manipulative(s)} & \textbf{Explanation}\\ {\sc
  Time Traveler} & \includegraphics[width=1.2em]{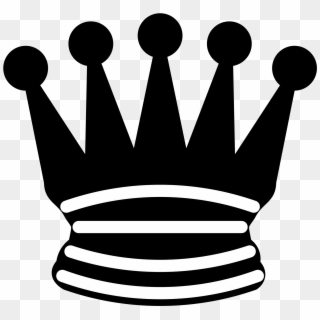} & In chess, the Queen can move linearly
in any direction: forward, backward, and diagonally.  Similarly, the
{\sc Time Traveler} role ‘moves’ both backwards and forwards in time,
and also explores the conditions that appear at those points in
time.\\

{\sc  Analyst} & \includegraphics[width=1.2em]{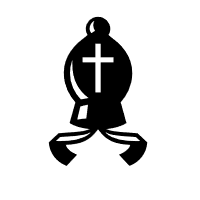}, \includegraphics[width=1.2em]{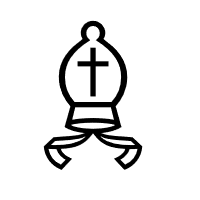} & In chess, there are two Bishops, both of which move diagonally, so that both are restricted to different colored squares. Here the {\sc Analyst} role divides its attention across two different spheres: articulations within the current challenge, and articulations of this challenge relative to other challenges.  \\

{\sc Wrinkler} &  \includegraphics[width=1.2em]{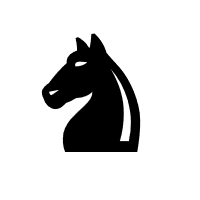} & In chess, the Knight moves in a skewed fashion: in order to go in one direction, it must also go a little bit in another direction.  The {\sc Wrinkler} role, similarly, looks at how a given strategy might go askew, due to unintended consequences or otherwise.\\

{\sc Linker} &  \includegraphics[width=1.2em]{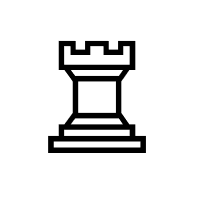} & In chess, the Rook moves any distance, as long as it goes in a straight line.  The {\sc Linker} role, similarly, can record a link between any two related concepts.  Since all concepts are potentially related in some fashion, the {\sc Linker} focuses on making \emph{useful} connections.\\

{\sc Reflector} &  \includegraphics[width=1.2em]{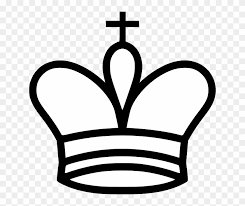} & In chess, capturing the King means the end of the game, so players are concerned throughout with any threat to their King.  Here, the {\sc Reflector} role senses how the discussion of a given scenario is progressing and when it would be good to draw it to a close and move on.\\

{\sc Stepper} &  \includegraphics[width=1.2em]{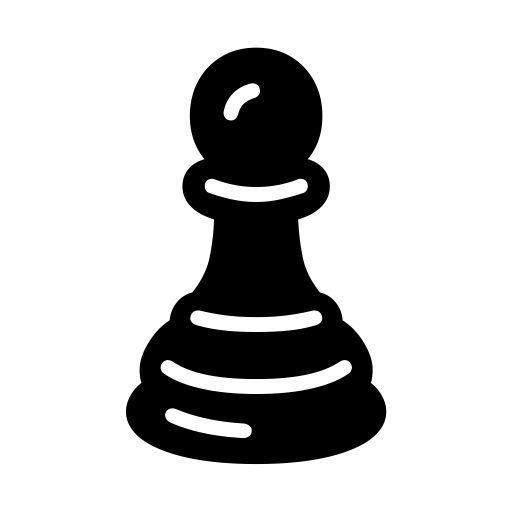} & In chess, Pawns move one space at a time (or two at the beginning of their movement).  Pawn are individually weak, but collectively, their placement is important for strategy.  The {\sc Stepper} role similarly describes the immediate next actions that should be taken at a point in time.
\end{tabular}
\caption{Mnemonic for manipulatives based on the chess set\label{mnemonic-for-manipulatives}}
\end{table}


\subsection{Intermediate artifacts}

\begin{figure}
  \includegraphics[width=\textwidth]{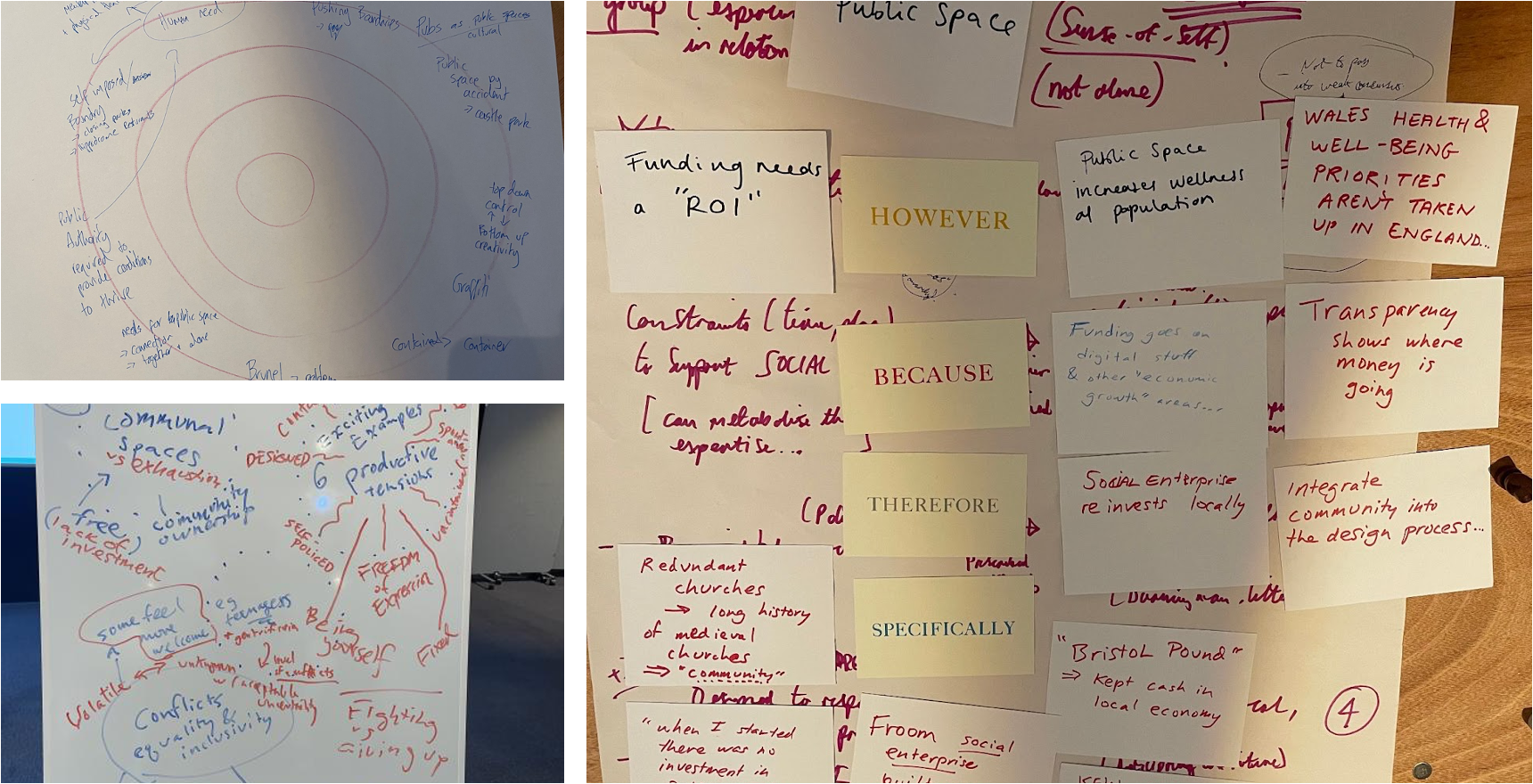}
  \caption{Use of diagrams and manipulatives to create {\sc Meaning
      Map}s and new patterns. In the upper-left
    photo, the CLA layers are mapped to concentric circles; from
    outside to center: litany, system, worldview, myth.  The {\sc
      Share Back} pattern was used to collect core themes from groups
    working separately, conceptualized here as the basis of a shared
    myth, comprising a {\sc Meaning Map} that pulls together the
    themes from small group discussions (lower left).  {\sc Pattern
      Language Components} were then used to sketch solution
    strategies to key problems and concerns, e.g., {\sc Funding of
      Public Space} (right). \label{ExampleParticipantPattern}}
\end{figure}

Intermediate artifacts generated within the workshop included mindmaps
created by participants on paper.  We used a graphical template
following the outline of the Causal Layered Analysis layers to explain
the workshop’s overall workflow, and also asked participants to use a
version of this diagram as a “grid” for note-taking within Phase I, to
encourage them to work from their observations to the core underlying
themes and issues.  Participants then clarified these core themes in a
share-back process, and in Phase II, developed them further in the
form of shared future stories, outlining paths to action.  Photos in
Figure \ref{ExampleParticipantPattern} show the movement from:
\begin{itemize}
  \item Initial sketching at the start of Phase I, beginning at the
    \emph{litany} level, within small groups, on to:
  \item A collection of themes shared across groups at the end of
    Phase I, to create a {\sc Meaning Map} which, in CLA terms, is
    intended to bring everyone into a shared \emph{myth} layer, on to:
  \item Phase II, using {\sc Pattern Language Components} to identify
    both general and specific possibilities for action around a theme.
\end{itemize}

\subsection{Output patterns} Participants created several patterns by
making use of the {\sc Pattern Language Components} and {\sc
  Functional Roles}.  For example, the right-most image in Figure
\ref{ExampleParticipantPattern} shows the participants’ articulation
of a {\sc Funding of Public Space} pattern (created with help from a
workshop facilitator).  Using the {\sc Pattern Language Components}
‘HOWEVER’, ‘BECAUSE’, ‘THEREFORE’, and ‘SPECIFICALLY’ made it easy to
write down structured candidate patterns (i.e., outlining the typical
Context/Problem/Solution triad, with the addition of ‘Next Steps’).
In Appendix \ref{catalogue}, these candidate patterns are
re-compressed into a succinct textual statement of the core idea of
the pattern.

Although in this workshop we were successful identifying in “paths to
action”,\footnote{\url{https://www.eventbrite.co.uk/e/public-space-for-public-health-a-call-to-action-tickets-492522286417}}
the main limitation was in regard to follow-through.  We had initially
planned to adapt Org Roam or some other similar tool into a wiki that
participants could use after the workshop to keep track of patterns
and next steps, recording progress, any blockers or ‘reverse salients’
\cite{hughes1993networks}, and any evidence for or against the
patterns, as well as their elaboration.  Our view is that other
technologies such as the Federated Wiki \cite{cunningham2013a} would
have required too much training time to be used for the purpose we had
in mind.  In the end, we simply didn’t deploy an end-user-writable
software platform.

\subsection{Evaluation}
\begin{description}
    \item[Methods] We expanded the system of manipulatives to include
      CLA diagrams, a {\sc Share Back} process, and explicit {\sc
        Pattern Language Components}.  Elicitation of structured
      patterns required active facilitator involvement.
    \item[Results] Mapping user contributions by offsite facilitators
      was both intensive, and rather subjective in nature.  This could
      be improved by reworking the mapping tools so that they could be
      used in real time by workshop attendees.
    \item[Interpretation] Firstly, the workshop was developed as paid
      consulting work, connecting the overall process overtly to the
      themes of the \textbf{market}.  The results of our offsite work
      could be viewed by attendees, but were not shared in a form
      where they could be directly extended.  Nevertheless, within the
      workshop itself, we saw attendees successfully adopt pattern
      methods, and use them to talk about potential futures.  This
      suggests the possibility for benefits accruing at the
      \textbf{state} level, with the potential for an Open Future
      Design workshop to function as a new kind of forum or senate, by
      analogy to developing practice around citizen juries.
\end{description}



\section{Case Study 3: Open Research Futures}\label{third-case-study}

This workshop was developed as an “Away Day” for faculty and staff
members at Oxford Brookes University.  The aim of the workshop is to
elaborate the institution’s open research strategy relative to its
existing organizational strategy.  Methodologically, this workshop
builds on a pre-seeded Org Roam network of interlinked themes and an
additional activity that enlists attendees in taking concrete actions
on the identified next steps.  The itinerary for this workshop adopted
the language “experts to citizens”, “citizens to action” from the
previous workshop.  These phases mirror the {\sc Dérive Comix}—{\sc
  Meaning Map}—{\sc Reinfuse Expertise} structure introduced in the
first case study; in CLA terms, the overall effect is a journey from
\emph{litany-to-myth} and then \emph{back-from-myth-to-litany}, with
the new litany taking the form of potential actions, or more
specifically, potential future headlines describing the actions.



\subsection{Intermediate artifacts}

\begin{figure}[h]
\begin{tabular}{c@{\hspace{3em}}c}
\includegraphics[trim={.2cm 0cm 0cm 2cm},clip=true,width=.4\textwidth]{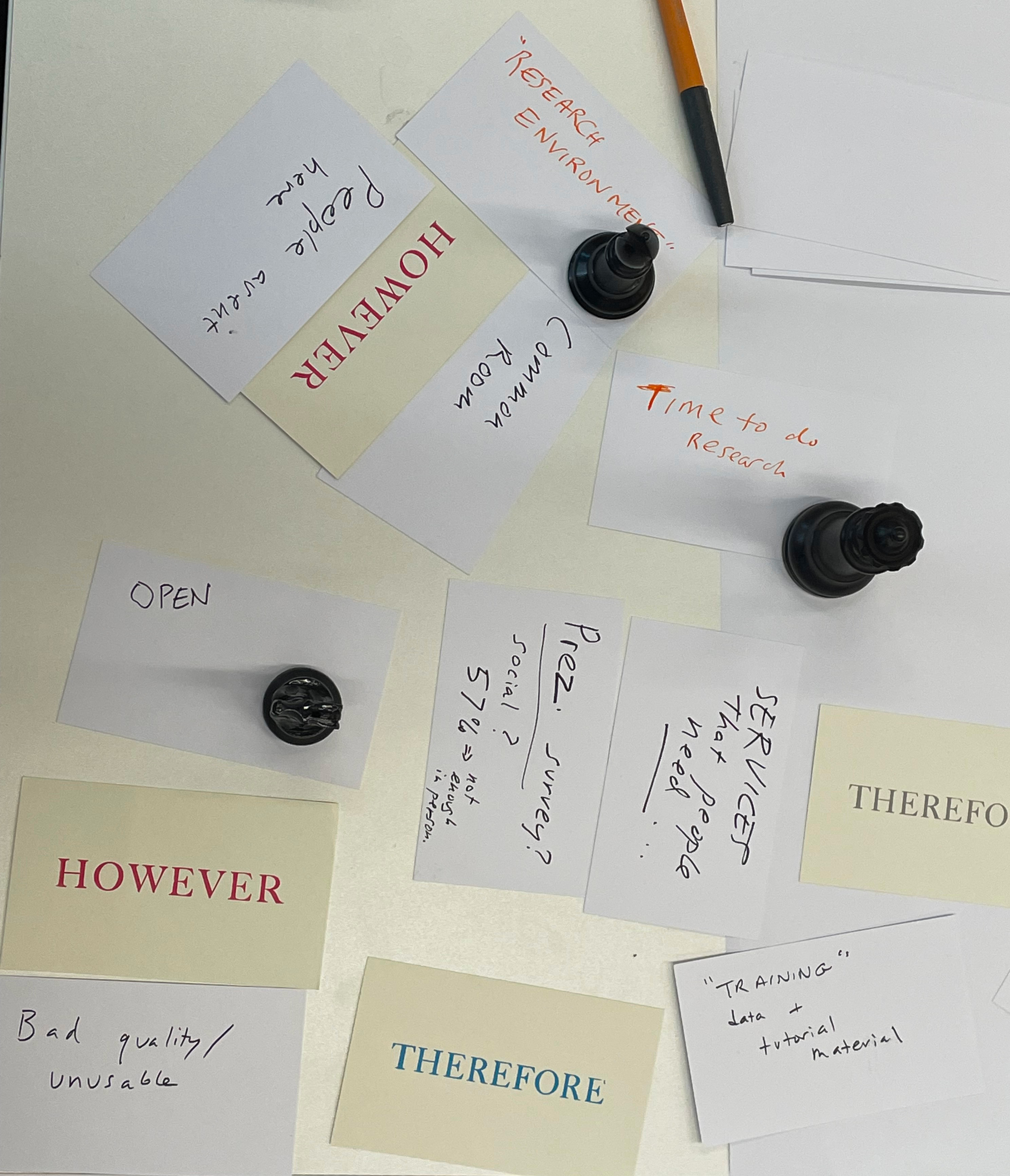}  &
\includegraphics[trim={0cm 1.4cm 0cm 2cm},clip=true,width=.4\textwidth]{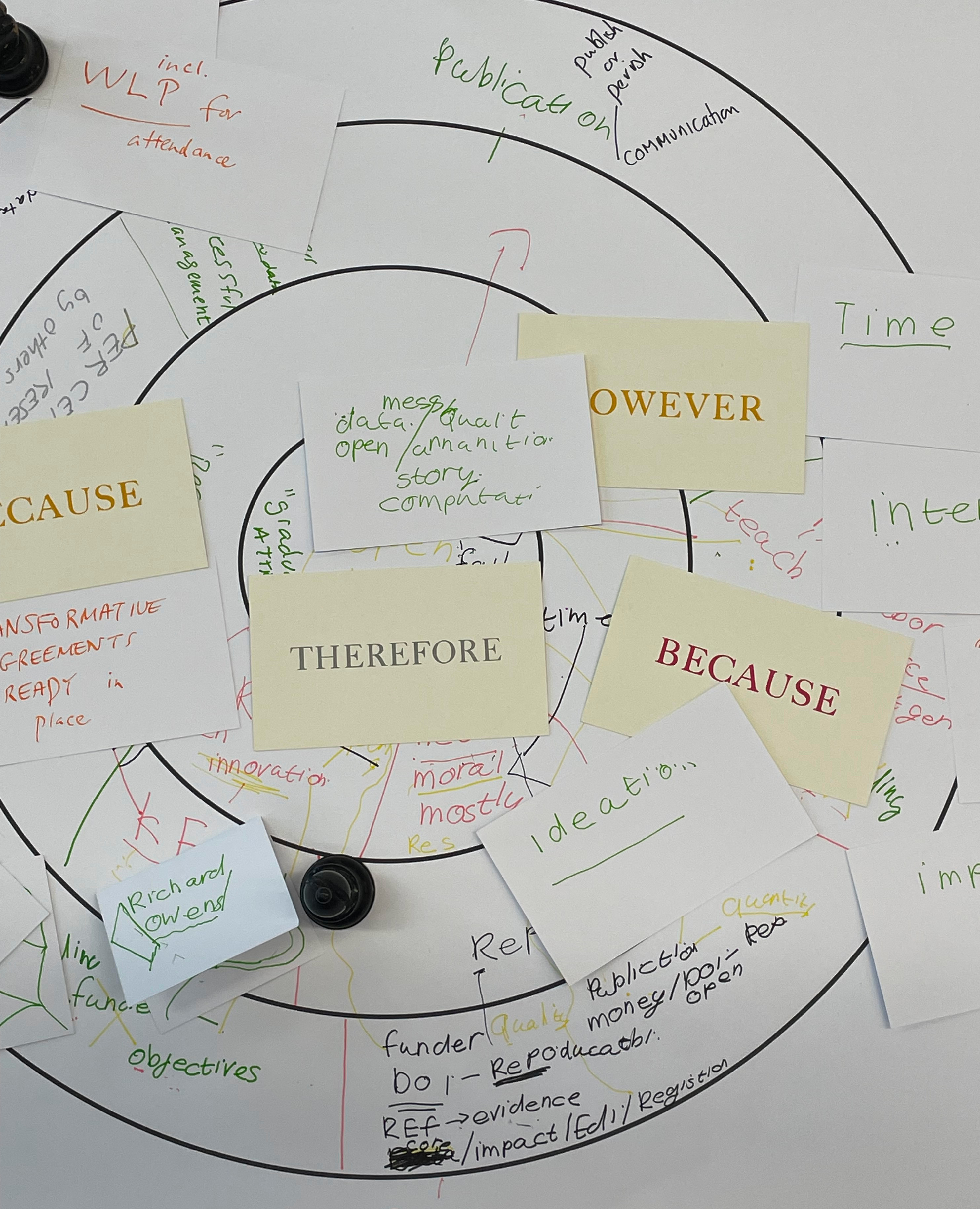}
\end{tabular}
\caption{The {\sc Pattern Language Components} were used organically within the workshop (cf.~\candidate{Destructure Patterns}).\label{brookes-workshop}}
\end{figure}

Figure \ref{brookes-workshop} shows how the {\sc Pattern Language
  Components} were used in the second phase of the workshop, building
on a CLA-based discussion that developed the {\sc Meaning Map} in the
workshop’s first phase (background, right).  In this case, the
keywords and manipulatives corresponding to {\sc Functional Roles}
(cf. Table \ref{mnemonic-for-manipulatives}) were not used to spell
out entire draft patterns (as in Figure
\ref{ExampleParticipantPattern}), but rather, to generate a network of
relations.

\begin{figure}[h]
  \includegraphics[width=\textwidth,trim={0 0 0 2cm},clip=true]{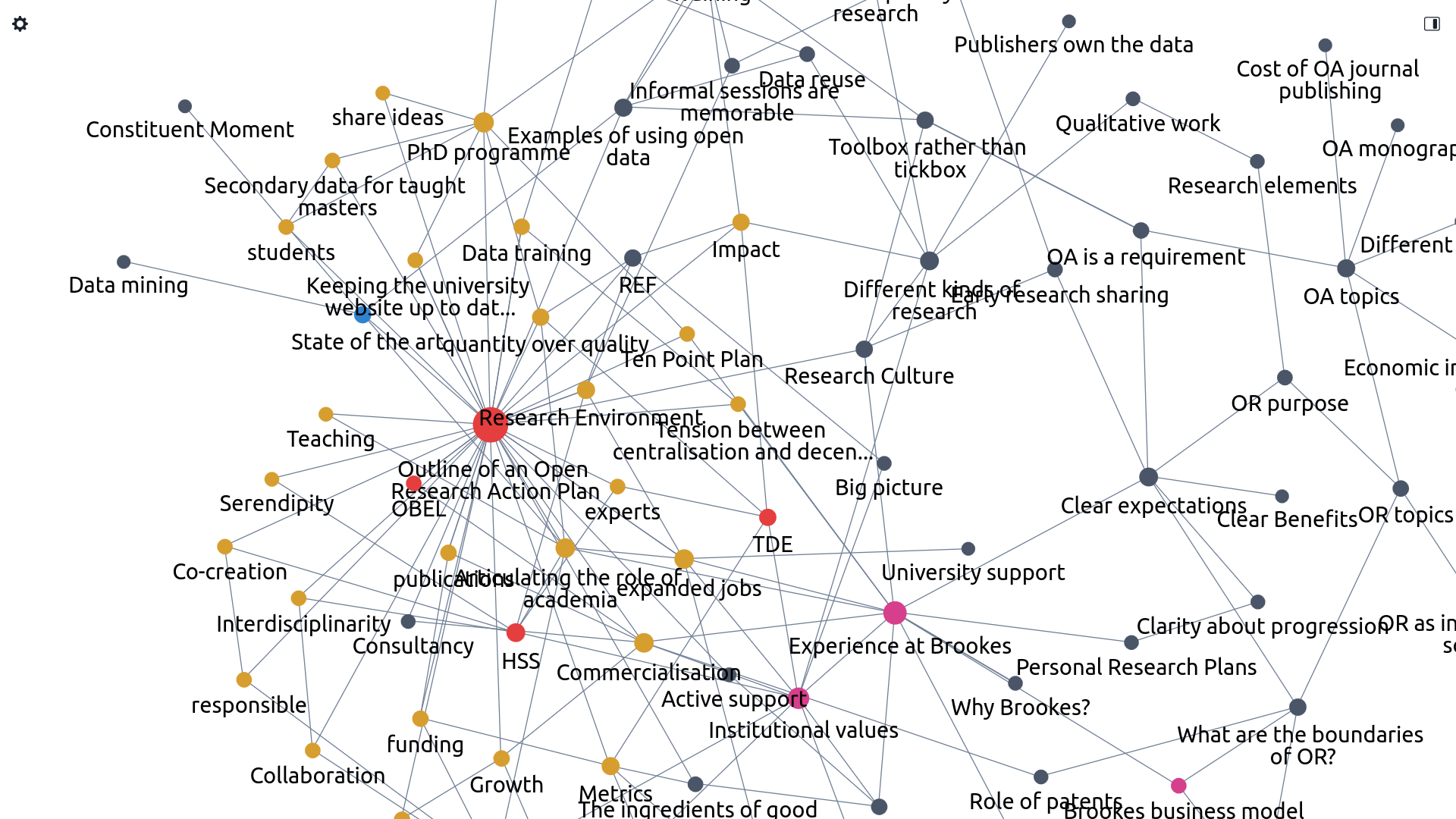}
  \caption{Screenshot of Org Roam UI, showing the development process
    leading to a draft Open Research Action Plan (ORAP).  Color-coding
    is: (Gray) Background themes and concepts based on interviews
    (cf. {\sc Do Your Research}); (Purple) Selected themes from that
    background material which became the focal themes in the workshop;
    (Yellow) Workshop themes and concepts; (Red) Key points of
    organization for workshop themes, including discussions per
    faculty as suggested by {\sc Structure Conversations}. The node
    “Outline of an Open Research Action Plan” includes the ORAP,
    instantiated here as a bullet-point outline, with links to all 
    the workshop outputs.
    \label{Org-Roam-Screenshot}}
\end{figure}

Figure \ref{Org-Roam-Screenshot} shows how material was then drawn
together, using Org Roam to analyze the workshop themes (per the {\sc
  Linker} pattern), further elaborating the {\sc Meaning Map}.
Contents of the paper-based diagrams, like those in Figure
\ref{brookes-workshop}, were condensed and edited in the digital
notes, rather than represented verbatim.  For example, the idea of
reintroducing a ``common room'' (Figure \ref{brookes-workshop}, left)
is folded into the ``Research Environment'' node, along with the
challenge of holding conversations when “people aren’t here” in
person.  The digital notes also provided an opportunity for
corrections, consolidation, and synthesis of links which hadn’t been
spelled out directly with the manipulatives. E.g., ``Richard Owens''
[sic] (Figure \ref{brookes-workshop}, right) is represented in the
digital notes as \emph{Richard Owen}, key proponent of responsible
innovation \cite{Owen2013}, within the ``responsible'' node.

The overall process illustrated in Figure \ref{Org-Roam-Screenshot}
moves in the opposite direction of \emph{Figures 3-7} of
\citet{iba2016pattern}, insofar as those figures complexify a tree as
a graph.  Here, we move from an interlinked graph of topics to a
summary map in tree form, represented here by the Outline of an Open
Research Action Plan node.


In brief, the organic, playful, interactions within the workshop were
useful in creating the more formal output (a draft policy document)
precisely because these interactions were informed by a suitable
metalanguage, including the apparatus of Causal Layered Analysis, the
{\sc Pattern Language Components}, the {\sc Functional Roles}, the
{\sc Meaning Map}, and other patterns described so far in the paper.
Counterfactually, we could have written up an action plan based solely
on desk research—e.g., writing a synthetic plan based on identified
organizational values “Success, Openness, Learning by doing,
Adaptability and Creativity, and Equal opportunity\ldots\ SOLACE”
\cite{corneli-varuf}.  However, that level of analysis only provides a
“mental picture”, in the terminology of \citet{alexander1964notes}.  A
hypothetical Open Research Action Plan based on stakeholder interviews
and the “SOLACE” concept, omitting the workshop experience, might
express the university’s stated values, but it would not draw on the
same richness of meaning, nor the same level of practical detail.



\subsection{Evaluation}
\begin{description}
    \item[Methods] We used the same manipulatives as in the previous
      case study, but this time with less structure.  This may suggest
      a level of reproducibility between that of generically useful
      PostIt\textsuperscript{\textregistered} notes and
      highly-structured design processes.
    \item[Results] The results were first collected in a graphical
      form, and then presented using a standard text template.  The
      pattern components were used to structure the map, though that
      process was still largely subjective.  Algorithmic additions to
      the methodology could enhance reproducibility, for example, by
      using the pre-existing template and a clustering algorithm we
      could (in principle) make the synthesis phase entirely
      reproducible.
    \item[Interpretation] The collated material was shared with a new
      working party that intends to create a new organizational
      strategy.  While the working party may ultimately develop a
      different interpretation of the role of open research from the
      picture created by workshop attendees, they will benefit from
      having seen that that picture.  This suggests a range of
      benefits accruing at the \textbf{market} level, both in
      the internal ideas marketplace, as workshop attendees will have
      their voice heard within the organization, and also at the
      organization level, as the university articulates its position
      within the sector in a way that incorporates on-the-ground
      realities.  As alluded to in the previous case study, a system for updating
      and maintaining diagrams like Figure \ref{Org-Roam-Screenshot}
      in real time would have uses in a form of “citizen science”.
      That suggests a technology-enhanced information ecosystem, with
      benefits at the \textbf{state} level.  However, from an
      organizational strategy perspective, the distilled bullet-point
      outline was an essential “product”, and the map itself contains
      too much information for immediate decision-making needs.
\end{description}

\section{Case Study 4: CIS 9590, Information Systems Development Project} \label{cis-case-study}

\subsection{Introduction to the course from the instructor, Mary Tedeschi}

CIS 9590 is Information Technology Project Design and Management is the  ``Computer Information Systems'' (CIS) capstone project course for the CIS major wherein the students will apply concepts and techniques from prior course work, to design, develop, and create an implementable application for a working information system of an actual business. It also focuses on the design and management of systems to meet the increased need for information within an enterprise. The course exposes students to the fundamentals of IT project management required for the successful implementation of IT-based systems. The course presents tools and technologies for project definition, work breakdown, estimating, planning and scheduling resources as well as monitoring and control of project execution. Students utilize knowledge gained from prior coursework, and work in groups to design and manage an Information Technology project.  During my first semester, Spring 2020, teaching with the students using whatever development tools they were familiar with, I noticed this to be a problem so with this knowledge I changed the course to require the use of Intel One API.  This did not get implemented until Fall 2021.  I actually taught the course three times before requiring students to use the same software tool uniformly.  The course was a 3-hour course, first face-to-face.  Then synchronous online only.  In Fall 2021 we changed to 75 minutes in person and online (hybrid).  Students had to self-teach Intel One API with the use of tutorials and a buddy system.  The students seemed to have the necessary skills to learn enough of the software to create an implementable application.  This semester, Spring 2023, the students really seemed to lack the coding skills.

\subsection{Our use of “Patterns of Patterns” within the course, by guest lecturers Raymond Puzio, Joe Corneli, and Charlie Danoff}
\label{pop1-in-cis9590}
Mary assigned our paper “Patterns of Patterns”
\cite{patterns-of-patterns-i} as a focal text with three successive
cohorts of CIS 9590 students.  The course syllabus is focused on
developing group projects with a computer programming component.  Our
hope was that the topics in the paper would inspire them with new
ideas about design and collaboration.  We focus primarily on the
latest iteration of the course (Spring 2023), in which we made the
most explicit use of the methods described above.  Figure
\ref{cis-9590-anticipations} shows some of our anticipations of the
relevant concerns that apply in this context.  Each semester, students
asked many thoughtful questions about the paper; each cohort also
produced their own collaboratively-written response to the paper,
engaging the original paper in depth; and in the latest run, we
offered some in-class exercises based on the workshop methods
described above.  Reading their written responses showed that the
students had not only understood the main ideas of our paper, but
added to them.  In effect, they created alternative imaginaries for
the paper’s history and future.  For instance, in their 2022 ‘case
study’, they generated a “Recommendation and Implementation Plan”
which proposed specific actions which a group could take based on our
ideas; and, in 2023, the students produced a slide presentation based
upon our paper, exploring its relationship to themes such as “emerging
technology”.  It is worth highlighting that while our paper did touch
briefly on the theme of emerging technology, the students considerably
elevated the importance of that theme in their feedback.

\subsection{Experience report by CIS 9590 student, Kajol Khetan}\label{sec:kajol-report}

The PLACARD method emphasizes understanding the context, selecting an
appropriate language or languages for thinking about problems in that
context, and taking relevant actions to guide the development process.
In our project work, we adapted the CLA component of this framework by
identifying the layers relevant to our chosen problem, namely, to
create a website that allows users to discover nearby coffeeshops
easily.  We identified the following layers for our analysis (each
with several facets): User Interface, Functionality, Data Flow,
Infrastructure, and External Factors.


CLA guides us to look for \emph{causal relationships} operating within
and between these layers.  In particular:
\begin{itemize}
  \item Changes in the User Interface layer influenced user engagement
    and ease of interaction.
  \item Adjustments in Functionality impacted the overall user
  experience and satisfaction.
  \item Data Flow optimizations directly affected the accuracy and
    relevance of coffee shop information.
  \item Infrastructure decisions impacted website performance and
    responsiveness.
  \item External factors influenced design choices and user
    satisfaction.
\end{itemize}

Our adaptation of Causal Layered Analysis led to a comprehensive
understanding of the website's components and their interdependencies,
and facilitated a structured approach to development.  The nearby
coffee shops website was successfully developed and deployed.  Users
can input their location and receive a map display of nearby coffee
shops, along with relevant information such as ratings, reviews, and
operating hours.  Our group’s process could be condensed into an
overall proto-pattern, \candidate{Adapt Layers as Needed}.

\subsection{Experience report by CIS 9590 student, Manvinder Singh}\label{sec:manny-report}

During my final project class led by Professor Mary Tedeschi, the
research paper “Pattern of Patterns” and its authors played a pivotal
role in shaping my project decisions.  The authors' active engagement
in our Zoom sessions provided invaluable insights into project layout,
helped me avoid common mistakes, and encouraged a scalable and
adaptable approach.  Their active engagement and promises to come
again in the following weeks to see progress on everyone’s individual
project kept the butterflies, nervousness, and willingness to deliver
all alive at the same time.  The guest lecturers’ contributions,
together with the way we adopted their paper, suggests several
proto-patterns which could be used to structure future versions of the
course: \candidate{Engagement and Guidance}, \candidate{Avoiding
  Mistakes}, and \candidate{Scaling and Adaptability}.




\begin{figure}[h]
  \begin{tabular}{cc}
    \includegraphics[width=.43\textwidth]{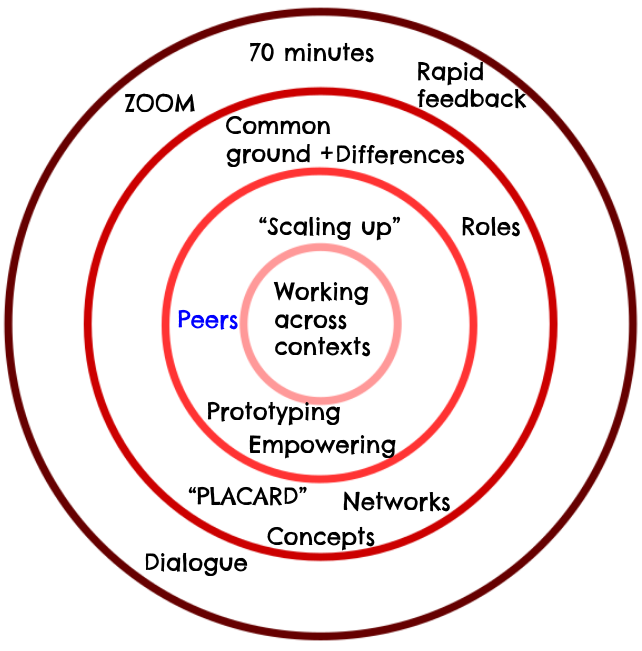} &
    \includegraphics[width=.45\textwidth]{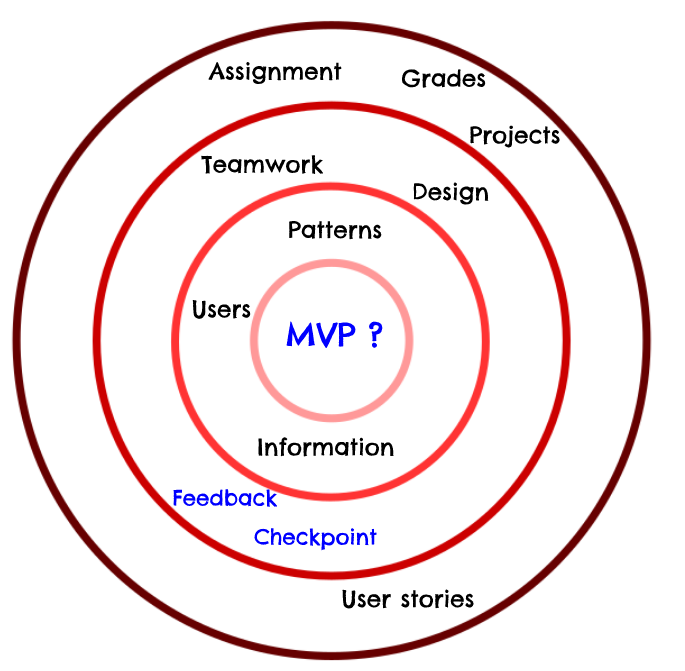}
  \end{tabular}
\caption{Diagrams created before our first session with the CIS 9590 students, inspired by Causal Layered Analysis.  The diagrams describe our working context as guests in CIS 9590 (left), and our initial understanding of the students’ working context (right).\label{cis-9590-anticipations}}
\end{figure}



\subsection{Evaluation}
\begin{description}
    \item[Methods] Guest-lecturing and receiving written feedback on
      our paper were directly replicable across cohorts.  The workshop
      methods introduced to the latest cohort adapted reasonably well
      to the online/hybrid classroom, though another round of
      prototyping could improve them.
    \item[Results] The student experience reports suggest a strong
      uptake of the methods we introduced.  With Mary, we discussed
      peeragogy-inspired learning design suggestions that could be
      used in future classes.
    \item[Interpretation] The experience went well enough that we were
      invited back several times, and, following the latest sessions,
      invited to continue the conversation with other instructors in
      the New York region (at PACE and Baruch).  Students reported
      improvements to their product design and development processes,
      with benefits at the \textbf{market} level.  We also see the
      potential for benefits accruing at the \textbf{commons} level,
      working across contexts by openly sharing and interlinking {\sc
        Meaning Map}s like those depicted in Figure
      \ref{cis-9590-anticipations}. 
\end{description}

\section{Case Study 5: PLoP 2023 Shepherding and Writers’ Workshop} \label{plop-case-study}

Our account would be incomplete without an overview of the way our
thinking, and this written presentation, continued to evolve in the
Pattern Languages of Programs 2023 context.  As in Case Study 1, at
the Writers’ Workshop, we were primarily asking for feedback.  As in
Case Study 4, participants had an entire paper to wrap their heads
around.  Our a result of the discussions with our shepherd and
workshop attendees, we clarified our use of proto-patterns, ultimately
focusing our presentation of the paper by moving the patterns and
proto-patterns to the appendix.  We rewrote the introductory sections
to make it clearer that our focus is on the process of evolving
patterns, rather than the presentation of polished patterns (more
typical at PLoP).

\subsection{Input patterns}
We put our pattern catalogue and surrounding narrative forward for
scrutiny and discussion.  The process was governed by pre-existing
patterns for paper shepherding and Writers’ Workshops.  We continued
to use patterns like {\sc Structure Conversations} after the workshop,
building in several rounds of internal peer review.

\subsection{Intermediate artifacts}
\begin{figure}[h!]

{\scshape
\clock{6}{10}{45}
}

\vspace{-.4cm}
\caption{Relationships between the component methods of PLACARD and the patterns introduced in this paper.    For example, the sensory dimension, PAR, serves to  “Identify themes”.  This process can be implemented through the {\sc Dérive Comix} and {\sc Share Back} patterns; further nuance can be added with the {\sc Context setting} pattern along with the {\bfseries\scshape Time Traveler} and {\bfseries\scshape Reflector} roles; and still more articulation can be added with the {\sc Do Your Research} pattern.   The three layers of patterns correspond to our workshops [1], [2], and [3], and roughly align with the CLA \emph{litany}, \emph{system}, and \emph{worldview} layers. Proto-patterns that were generated at any workshop or in subsequent analysis live at the myth level, denoted by {\normalsize \includegraphics[width=1.2em]{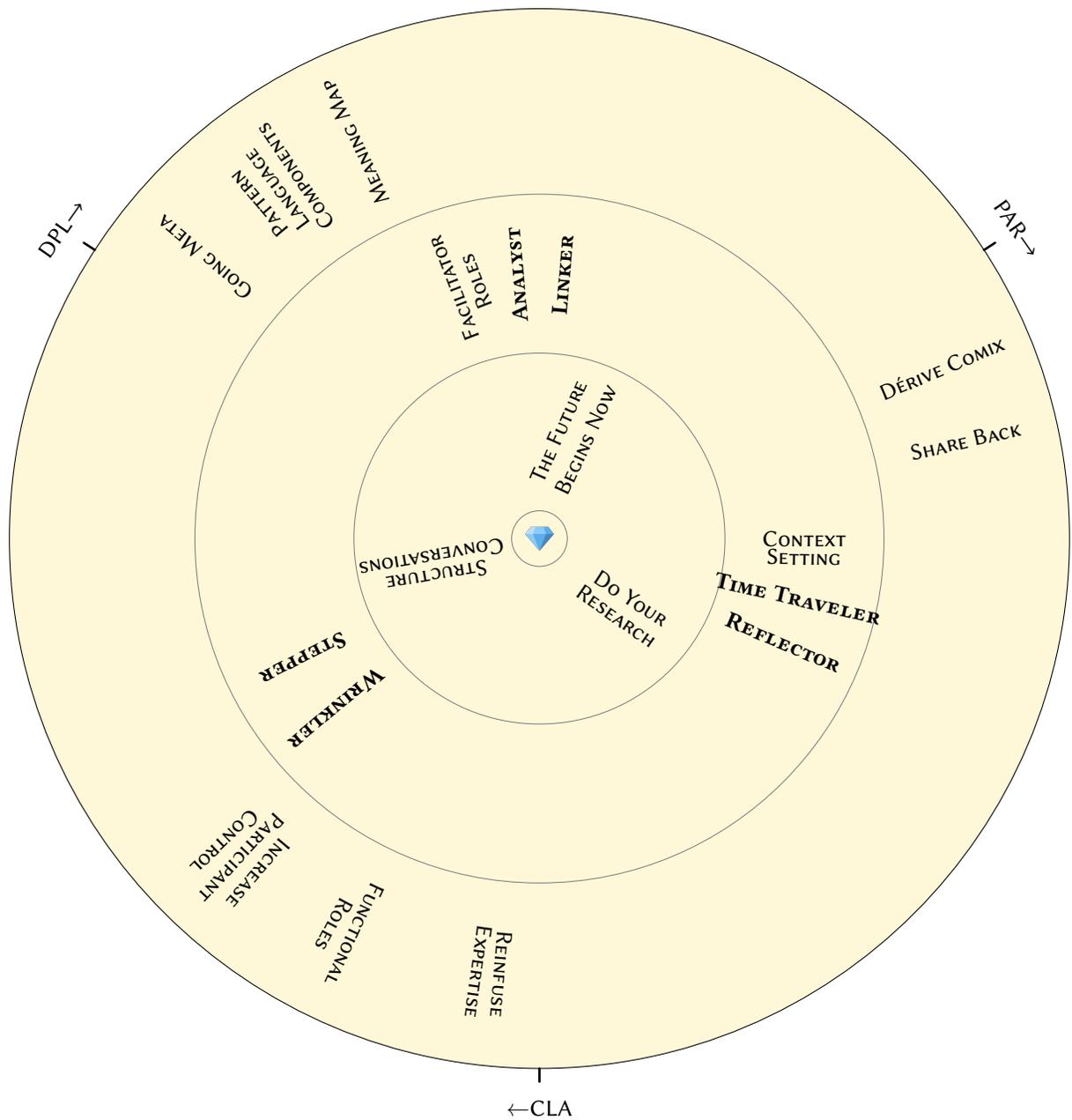}}.
  \label{alex-diagram}}
\end{figure}

We began by significantly shortening the paper, moving around 8000
words from our conference submission \cite{corneli2023patterns} into
an “outtakes” directory.  We restructured Table \ref{summary}, and
combined two earlier figures into Figure \ref{alex-diagram}.  In
particular, Figure \ref{alex-diagram} now shows both the abstract
workflow of pattern synthesis following the PLACARD method, and the
concrete process of pattern development that we followed.  These
correspond to outside-in and inside-out readings of the diagram.
Further significant revisions to the text were made.  In particular,
the evaluation sub-sections were added.

\subsection{Evaluation}
\begin{description}
    \item[Methods] The methods used at PLoP are well-established,
      having evolved into relatively firm traditions over 30 years.
      These were augmented by peeragogy patterns \cite{corneli2015a}
      and ideas from the current paper.
    \item[Results] Our paper received significant
      revisions focused on addressing the main takeaways from the
      workshop.  No new patterns were added, but the existing patterns were revised and the catalogue restructured.
    \item[Interpretation] PLoP methods are useful for polishing an
      existing set of patterns.  In and of themselves, PLoP workshops
      do not provide a complete recipe for creating useful pattern
      languages.  PLACARD and the associated patterns in this paper
      are therefore complementary to the existing PLoP methods.
      Having been through one complete cycle of research, writing, and
      revision since we presented “Patterns of Patterns”, we have now
      developed a complete methodic description for working with
      patterns in workshop settings.  We have touched on potential
      benefits across the four economic domains mentioned above.  The
      process of revision associated with PLoP helped us articulate
      these benefits in a cohesive manner, producing a more salient contribution
      to a knowledge \textbf{commons}.
\end{description}

\section{Discussion}\label{discussion}







\subsection{Related Work}\label{related-work}











We have taken a design-patterns-based approach to meta-research, using
design pattern methods to study design patterns.  Broad thematic areas
within meta-research include: “methods, reporting, reproducibility,
evaluation, and incentives (how to do, report, verify, correct, and
reward science)” \cite{Ioannidis2015}.
In this connection, Figure \ref{alex-diagram} has a theory-building
role, showing how the PLACARD pattern was expanded into further design
patterns which we developed across Case Studies 1-3.  Whereas common
problems in quantitative research are relatively well established
\cite{Munaf2017}, to date, the corresponding challenges and solutions
in qualitative research have been less well studied.  The diagram in
Figure \ref{alex-diagram} depicts both a research cycle for generating
new design patterns, which can now be run with variations, repeatedly,
and also a roadmap for deepening our collective understanding of and
facility with pattern-based inquiry.  \citet{HEIBERG2022104363} make
use of a similar diagrammatic approach to map evolving relationships
between technical and institutional concepts, superimposed upon a
‘radar plot’ which shows which concepts are central and which are
peripheral.

\FloatBarrier






\subsection{A remark on the Roles of Roles}\label{role-of-roles}



In developing the PLACARD pattern, we were inspired by the classical
neuroscientific notions of sensory, cognitive, and motor faculties.  The
Active Inference Framework (AIF) provides a more contemporary approach to similar concepts: see \citet{SMITH2022102632} for a
primer.  Here, we suggest that AIF and DPL
can support each other, with AIF concepts being given a practical
implementation using design pattern methods, and DPL receiving a
needed theoretical upgrade by way of Active Inference.

Both AIF and DPL have a broad scope of application, to a range of
complex systems.  DPL and AIF both make use of hierarchical
organization, e.g., \emph{A Pattern Language} cascades from patterns
for towns to buildings to construction; whereas AIF typically zooms
out, from simple to complex and reflective—from “I model the world” to
“we model the world” to “we model ourselves modelling the world”
\cite{Kirchhoff2018}.  Both AIF and DPL have a fundamentally
generative orientation.  Despite these similarities, AIF relies on
quantitative methods coming from statistics and theoretical physics, whereas DPL is primarily qualitative.  We wish
to draw attention to the way in which the participant roles outlined
above can operationalize key aspects of the Active Inference
Framework.

\begin{description}
\item[\normalfont{The} {\sc Time Traveler}] elaborates \emph{a prior
belief over states} and the \emph{likelihood of specific
observations}.
\item[\normalfont{The} {\sc Analyst}] elaborates a \emph{generative
model}, with the division between inward and outward articulations
  corresponding to \emph{internal} and \emph{external}-facing states
  (both with \emph{sensory} and \emph{active} components).
\item[\normalfont{The} {\sc Wrinkler}] elaborates a \emph{factor of
surprisal}.
\item[\normalfont{The} {\sc Stepper}] takes action to correct
  discrepancies between the generative model and the perceived state
  of the world, minimizing \emph{prediction error}.
\end{description}

Active Inference is currently being explored as a new paradigm for
artificial intelligence, with a focus on ecosystemic intelligence
\cite{friston2022designing,albarracin2023designing}.  In particular,
the AIF-inspired “hierarchical generative model capable of
self-access” from \citet{albarracin2023designing} arises through the
addition of reflective meta-layers, ascending from:
\smallskip

{
\renewcommand*{\arraystretch}{1.2}
\begin{tabular}{lllp{.37\textwidth}}
  &``\emph{What am I trying to do?}''&and&``\emph{What am I perceiving?}''\\
to:&``\emph{What am I paying attention to?}''&and&``\emph{What am I trying to pay attention to?}''\\
to:&``\emph{How aware am I of where my attention is?}''&and&``\emph{Am I trying to maintain awareness of my \phantom{X} attentional state?}''
\end{tabular}
}
\smallskip

This is similar to the \emph{litany-to-myth} direction of analysis
that we used in the first phase of our workshops.  In our workshop,
the roles were introduced in the \emph{myth-to-litany} phase, to
generate potential actions.  The socio-historical theory of
P.~R.~Sarkar, which informed the development of CLA, hinges on similar roles
(cf.~\citet{inayatullah1999situating,inayatullah1999transcending}).
The key point for our current purposes is that
{\sc Functional Roles} can help to bring about change across
levels and domains.

\subsection{Connection to other emerging computational technologies}
We reflect that pattern-based information processing tools
— commonly referred to as “artificial intelligence” — are being used
to model and influence a wide range of social and economic contexts.
Since AI and robotics routinely make use of concepts of cognitive science, this provides a common ground with the topics discussed here.  To flesh this out, AI techniques can be organized in nested layers, using CLA-style thinking, as we did with our patterns in Figure \ref{alex-diagram}.  Machine learning, neural networks and large language models correspond to the litany layer in the outer ring, since they work rather directly with raw data, detecting regularities (such as clusters and n-gram probabilities), and using these regularities to predict what comes next or to classify items.  In the next ring, we could place rule-based production systems and planners.  These correspond to the systems layer.  In the inner ring, we can place ontologies and knowledge representation languages, which formalize worldviews to make them amenable to computation.  For some contemporary thinkers, the innermost myth at the core is the possibility of building artificial general intelligence, though opinions vary about when and whether that will arrive, and alternative myths such as “AI empowers workers” have been proposed \cite{chiangNYer}.  We hope these observations will be of use in implementing design patterns computationally—perhaps starting with the patterns of patterns that we have collected—and/or in designing workflows for developing future AI systems.  

\FloatBarrier

\section{Conclusion}\label{conclusion}


Relative to previous work on the structure of patterns which used
a path-traversal metaphor to describe how patterns work
\cite{kohls2010a,kohls2011a}, a suitable metaphor for our contribution
is \emph{a wheel which rolls more smoothly when round ball bearings
are placed in the hub}.  Less metaphorically,
Richard Gabriel emphasized that the “patterns
and the social process for applying them are designed to produce
organic order through piecemeal growth” \cite{gabriel1996patterns}.
We given some further articulation to this point, and our case studies have
shown how such articulation can make work with patterns more
effective, efficient, and scalable.  In brief, our patterns of
patterns exemplify the latter part of the following description:

\begin{quote}
Technological progress is achieved through a dialectical relationship
between mediation (adaptation to the end terms: the path to be
travelled and the load to be carried) and autocorrelation, the
relation between the technical object and
itself. \cite{Simondon2005-pq}
\end{quote}

The primary limitation of this work is that it has been carried out
mostly at the “paper prototype” level.  Nevertheless, the process we
have presented reflects the main features we would expect from future
more technically developed platforms for distributed collaboration.
We described prototype-level distributed-working that was carried out
inside the workshops, organized via the {\sc Share Back} pattern; and
our workshop repeated with variation over time, allowing us to learn
from widely distributed teams.

Existing systems such as Alkemio\footnote{\url{https://alkem.io/}} are
being built with the need for collaboration across organizational
boundaries in mind, preferring “\emph{challenge-focused}”
collaboration.  A platform using DPLs to work \emph{across related
challenges} would take that idea even further.  Domain-level design
patterns outline potential new behaviors; improved support for the
process of gathering evidence that those behaviors do (or do not) in
fact work as intended is an ambitious but logical ramification of the
pattern method.  The {\sc Functional Roles} we’ve set forth here
provide an early informal articulation of the process, and the
connections to the Active Inference Framework that we alluded to in
Section \ref{role-of-roles} could scaffold further, more formal,
developments. 

Whatever underlying formalism is used, and independent of the specific
technologies that become part of future software implementations,
further work is needed to identify analogies between action arenas,
to highlight the ramifications of complex actions, to show predicted costs
and benefits, to search for tipping points that allow the effects of
change to reach across level boundaries, and to surface new questions for
consideration.  The overall need is summed up by \emph{Noema}
magazine’s editor-in-chief, Nathan Gardels: “It’s time for new
cooperative platforms that address irreducible
interdependence.”\footnote{\url{https://www.noemamag.com/from-globalization-to-a-planetary-mindset}}

\begin{figure}[h]
\begin{tabular}{cc}
  \raisebox{1.5cm}{\includegraphics[width=3in]{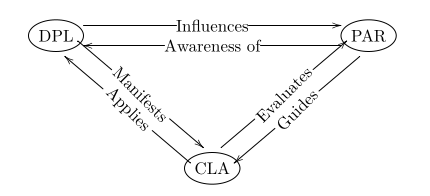}}&
  \includegraphics[trim={0cm 1.5cm 0cm 1.8cm},clip=true,width=3in]{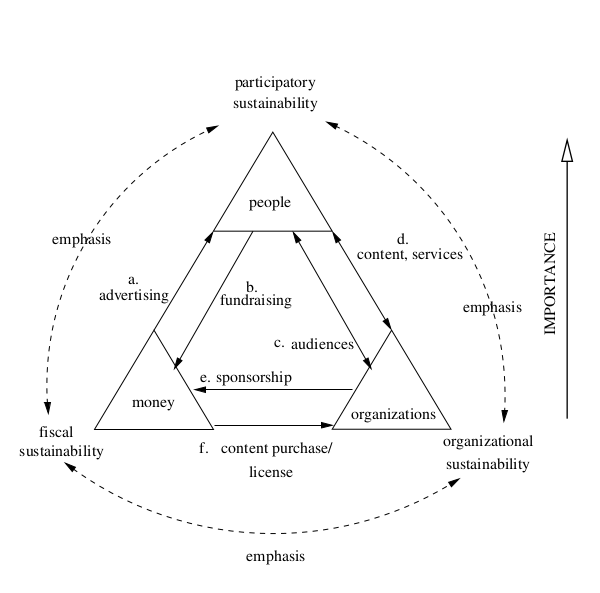}
\vspace{-.2cm}
\end{tabular}
\caption{The interrelated components of PLACARD (left) could scaffold
  a range of beneficial platform design principles (right, from
  \citet{krowne2003architecture}, used with permission).  \label{alex-to-aaron}}
\end{figure}

To illustrate how our work could help to address that need, Figure
\ref{alex-to-aaron} juxtaposes the relationships between the
components of PLACARD with a design diagram outlining the key sustainability features
abstracted from experiences with an early online community \cite{krowne2003architecture}.  If
PLACARD were used to design and build a system with features from 
the diagram at right, we might expect to see system-builders employing:
\begin{itemize}
\item DPLs related to money, organizations, and social interactions,
\item a CLA (or a collection of CLAs) that expresses the purpose(s) of the organization,
\item and ongoing PARs that help to keep things on track.
\end{itemize}

\begin{figure}[h]
\includegraphics[width=.6\textwidth]{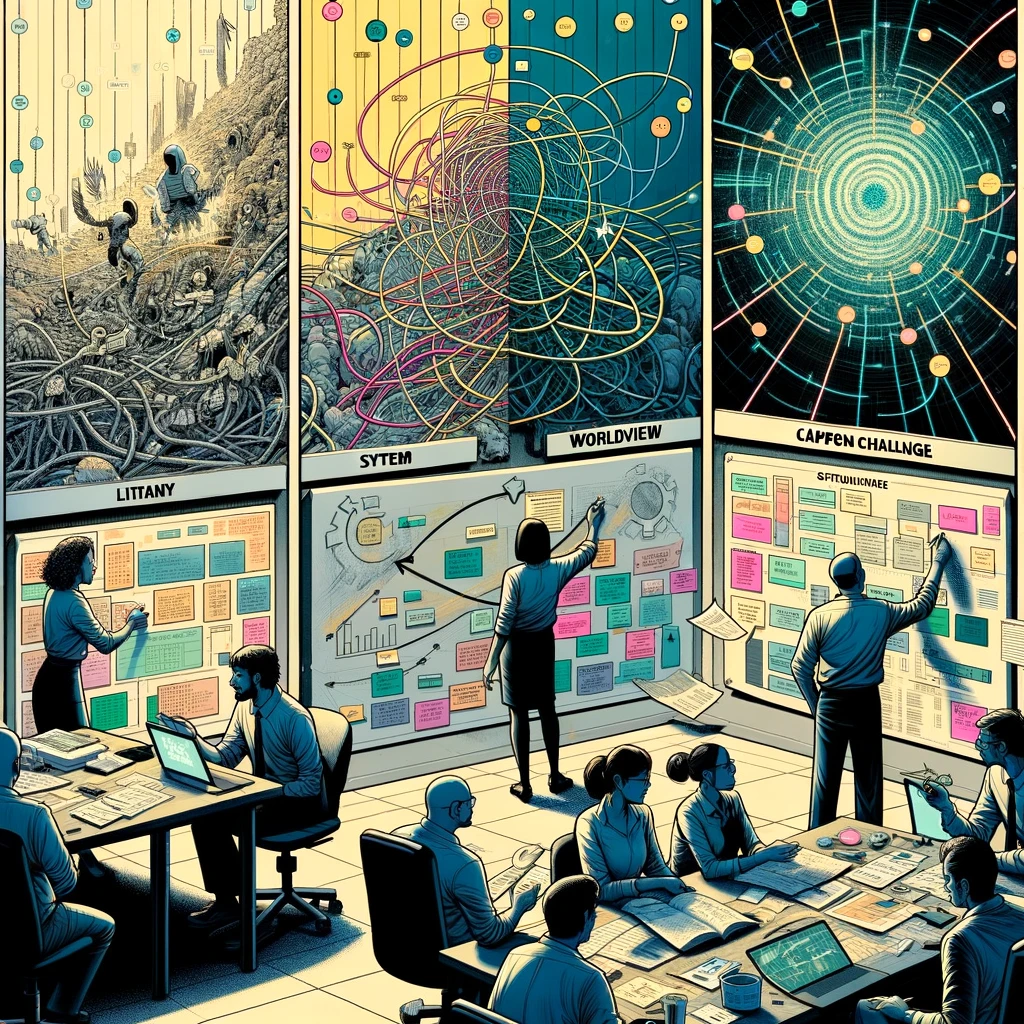}
\caption{A team facing a complex challenge depicted by a tangled web
  of lines and nodes indicating a complex problem.  The colorful orbs
  can be taken to represent other teams tackling related
  challenges. Illustrated by DALL-E.\label{team}}
\end{figure}
\clearpage

A community platform taking our collection patterns for “Identifying
themes”, “Organizing structure”, and “Making it actionable” as its
core content would be one possible instantiation.  This could be used
as a foundation for building further cooperative/collaborative
platforms.  One way to situate such
a platform would be as the “{\sc Going Meta}”
of contemporary efforts to model social and ecological systems
within the Doughnut Economics paradigm.\footnote{Downscaling the Doughnut: Data Portraits in action: A
collection of tools and useful examples for creating a Data Portrait
of Place, also known as City
Portrait. \url{https://doughnuteconomics.org/tools/92}}  The different local implementations of these ideas is a space to look for further design patterns.

Building on our remarks in the Discussion section, we suggest that the
existence of common patterns in social and artificial intelligence can inform the further development of the
theory and practice of social machines, building on the work of \citet{shadbolt2019theory}.  Patterns of patterns can
facilitate a unified approach to the
design and governance of such systems.  We have begun (as we mean to
carry on) by focusing on the development and articulation of
multi-purpose tools for thought.  We hope that future software systems
will better enable groups of collaborators and cross-group
collaborations.  Figure \ref{team} is dream-like AI-generated image,
showing what these interlinked collaborative settings might look like.

Given our assessment of the reproducibility of our methods, we would expect them to translate well to a variety of related research settings.  For example, in our comments on Figure \ref{alex-to-aaron}, we noted that PLACARD methods could inform platform design, without any commitment as to whether those systems would be digital or analog in nature.  In each case study, we evaluated our methods, with respect to whether they already have or could be extended to have benefits at the household, market, state, or commons level.   The result is a rigorously documented exploration of complex social interactions around relatively light-weight technologies for co-design and collaboration, which can serve as the jumping-off point for further stages of co-design and development.

Whether future work on patterns of patterns proceeds by developing computer programs that mine the space of potential patterns for those which support human thriving, or by building institutions that achieve something similar, or through some combination, developing the initial representations that support collaboration may be the hardest part.  Subject to the limitations of paper prototyping in hands-on workshops, and the pseudo-code-like nature of design patterns, our contributions can help scaffold those future efforts.

\section*{Acknowledgements}
We thank the participants in the workshops we hosted, and the CIS 9590
students who provided detailed feedback on both the prequel and the
current paper.  We wish to thank and acknowledge the funders who
supported our work.  The workshop described in Case Study 2 was
sponsored by a Springboard grant from the University of the West of
England; and the workshop described in Case Study 3 took place under
the auspices of the Research England REDF grant: “Growing and
embedding open research in institutional practice and culture”.  Kajol
Khetan’s contributions were supported by a Sidney and Laura Gilbert
Internship Award from Baruch College.  We would also like to thank our
PLoP 2023 shepherd Kiyoka Hayashi who gave detailed comments on
successive drafts of this write-up.  Charlotte Pierce also provided
helpful comments on a draft of the paper.  We gained further helpful
feedback at the PLoP 2023 Writer’s Workshop, and additional
inspiration from material presented by Laura Fortunato at the
Oxford-Berlin Autumn School on Open and Reproducible Research.
\textit{From Mary Tedeschi}: I would like to thank Pai-Chun Ma, Nanda
Kumar and Rudy Brown, who allow me to be creative in my classroom.

\appendix

\section*{Appendix: Pattern Catalogue} \label{catalogue}
\renewcommand{\thesubsection}{\Alph{subsection}}

We use this simple variation on the classical “Context/Problem/Solution” or “Context/Conflict/Solution” \cite{alexander1970a} pattern
template:

\smallskip
\begin{tabular}{lll}
\phantom{hellohellohello}&\textbf{Context} \ldots & {[}Summary of the working context{]}\\
\phantom{hellohellohello}&\textbf{If} \ldots\ \emph{BUT} \ldots & {[}A conflict, problem, or gap arising in this context{]}  \\
\phantom{hellohellohello}&\textbf{Then} \ldots & {[}Actions to take to resolve the conflict{]}
\end{tabular}
\smallskip

\noindent We sometimes add an \textbf{Example} where we feel that
would be helpful.  In this connection, it is worth highlighting that
{\sc Pattern Language Components} pattern covers similar ground with a
different lexicon:

\smallskip
\begin{tabular}{lll}
\phantom{hellohellohello}&\ldots\ \textbf{HOWEVER} \ldots & {[}A conflict, problem, or gap{]}\\
\phantom{hellohellohello}&\textbf{BECAUSE} \ldots & {[}A solution to a similar problem arising elsewhere{]}  \\
\phantom{hellohellohello}&\textbf{THEREFORE} \ldots & {[}High-level actions to adapt and try in this context{]} \\
\phantom{hellohellohello}&\textbf{SPECIFICALLY} \ldots & {[}Specific next steps to begin with{]} \\
\end{tabular}
\smallskip

\noindent Some of the patterns concern roles taken on by workshop
participants, and for these we use an adapted template:

\smallskip
\begin{tabular}{lll}
\phantom{hellohellohello}&\textbf{Context} & {[}\emph{As above.}{]}\\
\phantom{hellohellohello}&\textbf{Question}\ldots & {[}The thematic question that this role is concerned with{]}\\
\phantom{hellohellohello}&\textbf{Role}\ldots & {[}What purpose the role fulfils in the conversation{]}  \\
\end{tabular}
\smallskip

\noindent Some proto-patterns are included which do not have any
template sub-structure.  They are typically \emph{ideas for patterns}
which have not been elaborated.  Their titles are written in a
mixed-case italic serif font.  Three of our collected proto-patterns
were simplified adaptations of more elaborate pattern descriptions
created by workshop attendees (see for example Figure \ref{ExampleParticipantPattern}) and are marked with a “\includegraphics[width=1.2em]{gem}”.  In contrast to Table
\ref{summary} which outlines both patterns and proto-patterns in historical order of their development, for ease of reading, this appendix collects the proto-patterns at the end of each subsection.  It should be noted that
the assignment of a pattern a subsection is somewhat subjective, and even more so for proto-patterns.

\subsection{Identifying themes} \label{it-a}
\smallskip

\subsection*{DÉRIVE COMIX}

\textbf{Context} you want to develop some future scenarios to explore with a group.\newline
\textbf{If} the group has been identified BUT the members don’t know each other well yet, and accordingly each has their own separate experience, and the group has no concrete shared meanings;\newline
\textbf{Then} Gather data.  For example: go for a walk \cite{debord},  or just look out the window wherever you are.  (Alternatively, close your eyes and conduct a mental exploration of your selected theme: what do you see in “your mind’s eye?”) Document what you see.  Follow up by preparing your materials to share in a succinct fashion, e.g., as photos, a screenshot, slides, sketches, a ’zine, a map, or some PostIt\textsuperscript{\textregistered} notes.\newline
\textbf{Example} Variants of this pattern can be used to explore potential futures, as we observed in the PLoP 2023 workshop “Future-Self Immersion with using A Pattern Language for Nurturing an Exciting Life”, where participants “envision a future where they actively implement a particular pattern.”\footnote{\url{https://www.hillside.net/plop/2023/index.php?nav=program}}

\subsection*{SHARE BACK}

\textbf{Context} Members of sub-groups have shared their experiences with each other, and each sub-group has
developed a {\sc Meaning Map} \newline \textbf{If} we want to
establish a meaning map for the whole group BUT unstructured
interaction within the whole room is infeasible;\newline \textbf{Then}
individual groups should present key findings to the room; or,
alternatively, if the conversations are not yet at a natural stopping
point ask the groups to pause their conversations and listen in briefly
to the one of groups (in turn), who will continue their small-group
conversation.

\subsection*{CONTEXT SETTING}
\textbf{Context} A workshop or other working context has been convened.\newline
\textbf{If} the facilitators have ideas that they would like to explore with attendees BUT these ideas are not likely to be top of mind for attendees.\newline
\textbf{Then} do some context-setting, e.g., give a short talk about why people have been invited, and describe the hoped-for outcomes.\newline
\textbf{Example} Prior to the workshop described in Case Study 2, short videos on the themes ‘public space’ and ‘public health’ were elicited from attendees.  At the start of the workshop these were projected to the assembled audience.  This was accompanied by further remarks that juxtaposed what we planned to do at the workshop with previous work on “interactive documentary”\footnote{\url{http://i-docs.org/interactive-documentary-what-does-it-mean-and-why-does-it-matter/}}.

\subsection*{TIME TRAVELER \includegraphics[width=1.2em]{queen.jpeg}}

\textbf{Context} In an interaction that can be structured with {\sc Functional Roles}.\newline
\textbf{Question} \emph{What has happened in the past, what could
happen in the future?}\newline
\textbf{Role} To provide historical context and
anticipate alternate futures.

\subsection*{REFLECTOR \includegraphics[width=1.2em]{king.png}}

\textbf{Context} In an interaction which can be enabled by people taking on {\sc Facilitator Roles}.\newline
\textbf{Question} \emph{How is the scenario evolving?}\newline
\textbf{Role} To appraise each developing scenario, provide a format
for reflection (e.g., PAR, or, indeed, if time allows, the entirety of PLACARD), make a decision to continue, reset, or end.

\subsection*{DO YOUR RESEARCH}
\textbf{Context} Prior to beginning a formal workshop or other participatory research activity, facilitators may have time available that goes beyond what will be asked of participants in {\sc Dérive Comix}.\newline
\textbf{If} it will be possible to do participatory research in the workshop setting BUT the context outside of the workshop is potentially just as important.\newline \textbf{Then}
start doing the research in a more centralised way before inviting direct collaboration. Findings from an earlier research stage can be summarised and presented at the {\sc Context Setting} stage to give participants something to engage with.\newline
\textbf{Example} In Case Study 3, the pre-research phase included 1-to-1 interviews with around half of the invited participants, as well as internet research to find and explore related scenarios developed by others in the sector.\footnote{\url{https://royalsociety.org/topics-policy/projects/research-culture/changing-expectations/visions-of-2035/visions-of-2035-materials/}}

\subsection*{\candidate{Pilot to Anticipate}}

Invoking the {\sc Going Meta} pattern (see Subsection \ref{os-a},
below), we reflect that our strategy of \emph{piloting our workshop
methods} was how we choose to anticipate the issues likely to arise in
future iterations of the workshop.  Perhaps the future of anticipation
more widely will include the increased use of pilot schemes?  In
support of this possibility, \citet{unger2019imagination} have
advocated for the enthusiastic embrace of “Experimental government”.

\subsection*{\normalsize{\includegraphics[width=1.2em]{gem}} \candidate{Contested Space}}\label{pat:contested-space}

So-called public space doesn’t always feel welcoming to all members of the public.  It can be overrun with antisocial behavior.  It can feel exclusionary, or uninviting.  It can be the site of conflict.  Although the uses of public space are complex, each space need not support every use equally.

\subsection{Organizing structure} \label{os-a}
\smallskip

\subsection*{MEANING MAP}

\textbf{Context} We have collected images or other data describing
people’s worlds (see {\sc Dérive Comix}).\newline
\textbf{If} our intention is to distill
well-integrated shared understandings with the group BUT, so far,
everyone has been keeping most of their experience, knowledge, and
perspectives to themselves;\newline
\textbf{Then} talk together about the problems and opportunities that everyone sees in the data that has been gathered and shared, and document any connections you find.
You can return to the exploration activities of {\sc Dérive Comix} as needed.
Maybe some of the identified themes will start to cluster together.
Maybe everyone will have wildly different perspectives: that’s also entirely OK.
Either way, you can use these different viewpoints to bring everyone on the same page: just document them as part of the map.

\subsection*{PATTERN LANGUAGE COMPONENTS}

\textbf{Context} In a collaborative setting with people who are new to
design patterns.\newline \textbf{If} attendees are being invited to
create new design patterns that operationalize knowledge at the group
level BUT the typical framing language of DPLs — which have ‘conflict’
at the core — is not comfortable for participants (e.g., because a
‘problem’ or ‘conflict’ is seen as a bad thing);\newline \textbf{Then}
introduce and describe neutral keywords such as \emph{HOWEVER} (which
can variously be used to describe a gap, a conflict, an opportunity,
or even a simple juxtaposition of facts), \emph{BECAUSE} (to describe
a set of operating causes), \emph{THEREFORE} (to describe a rationale
based on related data), and \emph{SPECIFICALLY} (to describe next
steps), to help people build patterns piece by piece.

\subsection*{GOING META}\label{method1}
\label{sec:org958ff04}

\textbf{Context} In the course of working on a project
together.\newline \textbf{If} we find a \textit{gap} between our
ideals and our methods;\newline \textbf{Then} Try “going meta”, to
explore how the project’s methods can be applied to itself.\newline
\textbf{Example} In a community that usually focuses on anticipating
the future for others, try inviting members of the community to
anticipate the future of the community itself.

\subsection*{FACILITATOR ROLES}
\textbf{Context} Developing a collection of interrelated design
patterns.\newline \textbf{If} you are getting ideas from participants
who play {\sc Functional Roles} BUT the ideas aren’t all connected
with each other in a structured way; it’s hard to know when to move on
to another topic; and potentially, other obstacles arise.\newline \textbf{Then} introduce facilitator roles,
e.g., to help structure the work, and to decide when to keep the conversation moving or draw it to a close.

\subsection*{ANALYST \includegraphics[width=1.2em]{BBishop.png}, \includegraphics[width=1.2em]{WBishop.png}}

\textbf{Context} In an interaction that can be structured with {\sc Functional Roles}.\newline
\textbf{Question} \emph{What are the moving parts?}\newline
\textbf{Role 1} Consider the current challenge and all the components
of the potential solution (actors, resources, institutions). Identify
and orchestrate the dynamic network of these components.\newline
\textbf{Role 2} Consider the other challenges specified beyond the
current focus.  Identify and orchestrate the integration of these
components relevant to the present challenge.

\subsection*{LINKER \includegraphics[width=1.2em]{rook.png}}

\textbf{Context} In an interaction which can be enabled by people taking on {\sc Facilitator Roles}.\newline
\textbf{Question}
\emph{How do proposed scenarios build into patterns across layers, and how do they interact within the constellation of related concepts?}\newline
\textbf{Role.} Data wrangling as it comes in, providing visualization of patterns and interconnections.

\subsection*{STRUCTURE CONVERSATIONS}
\textbf{Context} Having convened a workshop or other participatory research activity.\newline
\textbf{If} unstructured discussions are likely to take lots of time BUT without yielding concrete benefits.\newline \textbf{Then}
structure the discussions around shared interests to move things forward more effectively.\newline
\textbf{Example} In the “Open Research Futures” workshop, we decided to group participants around tables according to the faculty where they were employed (or most closely aligned, in the case of university-level support staff).

\subsection*{\normalsize{\includegraphics[width=1.2em]{gem}} \candidate{Funding of Public Space}}\label{pat:funding-of-public-space}

Even though public space is known to increase wellness in the
population, well-being priorities that would lead to increased funding
for public space aren’t universally adopted.  In order to make the
benefits of such investment clear, increase transparency around
investments in public welfare, e.g., create a register of impacts of
local social enterprises.

\subsection*{\candidate{Destructure Patterns}}
The {\sc Pattern Language Components} need not be used to articulate
entire patterns: a less formal discussion can surface useful meanings.

\subsection*{\candidate{Adapt Layers as Needed}}
Layer-based analysis facilitates effective communication among team
members, enabling seamless collaboration, and aides both design and
implementation.  For this to work well, we need to select the right
layers.  In a complex change process, we might use CLA; in prototyping
project, relevant layers include the \emph{languages} and the
protocols describing an implementing suitable \emph{actions}.

\subsection*{\candidate{Avoiding Mistakes}}
Guidance from experienced developers can help avoid common project
development pitfalls.  Some useful methods include effective
documentation, regular testing, and thorough project planning.

\subsection*{\candidate{Scaling and Adaptability}}
By considering emergining technologies and incorporating modular
elements within a flexible framework, we can accomadate and adopt new
advancements.

\subsection{Making it actionable} \label{ma-a}
\smallskip

\subsection*{REINFUSE EXPERTISE}

\textbf{Context} a group wants to build a {\sc Meaning Map}.\newline
\textbf{If} everyone has experience as human being (and resident,
citizen, etc.) BUT they also have some experience as an expert which
is harder to share with non-experts;\newline \textbf{Then} begin by
removing expertise to get everyone on the same page, and subsequently
reinfuse expertise, to enable richer and more complex thinking.
\newline \textbf{Example} In preparing the first edition of the
\emph{Peeragogy Handbook}, we worked together informally until we had
buy-in from around 25 contributors together with a high-level outline
of the main themes we wanted to discuss.  This outline was then filled
in with individual chapters, most elaborating the specific experience
of one or two
co-authors.\footnote{\url{https://en.wikibooks.org/wiki/Peeragogy_Handbook}}

\subsection*{FUNCTIONAL ROLES}

\textbf{Context} When building a new set of design patterns.\newline
\textbf{If} you have ideas about the components of a pattern BUT the
pattern hasn’t been fully formed yet.\newline \textbf{Then} introduce
different perspectives to critique the pattern as it
develops.

\subsection*{WRINKLER \includegraphics[width=1.2em]{knight.png}}

\textbf{Context} In an interaction that can be structured with {\sc Functional Roles}.\newline
\textbf{Question} \emph{What could go wrong? }\newline
\textbf{Role.} Consider what might derail or counter
the proposed solution.  Each wrinkle can be assigned a level of
perturbation (from low to high).

\subsection*{STEPPER \includegraphics[width=1.2em]{pawn.png}}

\textbf{Context} In an interaction that can be structured with {\sc Functional Roles}.\newline
\textbf{Question} \emph{What should we do next?}\newline \textbf{Role}
Consider the discussion so far, and the various possibilities for
action that have arisen.  Decide which actions would be most useful or
informative, and devise a plan in place to carry them out.

\subsection*{THE FUTURE BEGINS NOW}
\textbf{Context} Having developed possible next steps in a discussion.\newline
\textbf{If} it appears that leaving the discussion without concrete commitments means concrete actions are less likely to take place.\newline \textbf{Then} take preliminary actions before leaving the discussion to create a sense of commitment and follow-through.\newline
\textbf{Example} One way to build commitment would be to ask people to develop and share a method for a small-scale experiment that they plan to carry out.

\subsection*{\candidate{Increase Participant Control}}

When organising a collaborative activity, participants should not
remain only an audience, or only deliver scripted lines (as was
reinforced by our Anticipation 2019 experience, see Section
\ref{methods}).  Give them increasing responsibility. 

\subsection*{\normalsize{\includegraphics[width=1.2em]{gem}} \candidate{Rebalance Social Services}}\label{pat:rebalance-social-services}

Welfare-related services should be supplied in balance with local needs, though they often are not. Can varied expertise be integrated in a similar way to the domain-specific skills practiced by Médicins Sans Frontièrs\footnote{\url{https://www.msf.org/}} to address complex local challenges?

\subsection*{\candidate{Structure Outputs}}
Having gathered themes from a participatory project, they may have
some explicit (e.g., because of how the information was gathered,
cf. {\sc Structure Conversations}).  Additional structure can be
created, if you link intermediate artefacts into a relevant template.

\subsection*{\candidate{Engagement and Guidance}}
Guidance from pattern experts can help create a collaborative learning
environment, allowing participants to gain deeper insights into
relevant concepts and methodologies, and help foster to innovative and
logically-coherent project approaches.



\renewcommand\bibname{References}
\renewcommand\refname{References}

\bibliographystyle{ACM-Reference-Format-Journals}
\bibliography{./main.bib}


\end{document}